\documentclass[conference,letterpaper]{IEEEtran}
\newcommand{\beq}{\begin{equation}}
\newcommand{\eeq}{\end{equation}}

\newcommand{\lsb}{\left[}
\newcommand{\rsb}{\right]}

\newcommand{\bmeta}{\boldsymbol{\eta}}

\def\adots{\mathinner{\mskip0mu\raise0pt\vbox{\kern7pt\hbox{.}}\mskip3mu
          \raise4pt\hbox{.}\mskip3mu\raise8pt\hbox{.}\mskip0mu}}

\newcommand{\bfH}{{\mathbf H}}

\newcommand{\diag}{\mbox{diag}}

\newcommand{\tr}{\mbox{tr}}

\newcommand{\blkdiag}{\mbox{blkdiag}}
\newcommand{\bmh}{\bfh}

\usepackage{bm}

\newcommand{\bmtheta}{{\bm \theta}}

\newcommand{\bmx}{{\bm x}}
\newcommand{\bmy}{{\bm y}}

\newcommand{\bmR}{{\bm R}}
\newcommand{\bmP}{{\bm P}}

\renewcommand{\bmh}{{\bm h}}

\newcommand{\bmS}{{\bf S}}

\newcommand{\bmA}{{\bf A}}

\newcommand{\bmn}{{\bm n}}

\newcommand{\bmI}{{\bf {I}}}

\newcommand{\bmzero}{{\bm 0}}

\newcommand{\bma}{{\bm a}}

\newcommand{\CRB}{\mbox{CRB}}
\newcommand{\FIM}{\mbox{FIM}}

\renewcommand{\bmeta}{\boldsymbol{\eta}}

\newcommand{\E}{\mbox{E}}

\newcommand{\bmb}{\mathbf{b}}
\newcommand{\bmr}{\mathbf{r}}
\newcommand{\bms}{\boldmath{s}}
\newcommand{\bmg}{\mathbf{g}}
\newcommand{\bmB}{\bf{B}}

\newcommand{\bmX}{{\mathbf X}}

\newcommand{\bmnu}{{\bm \nu}}
\newcommand{\hbmnu}{{\hat{\bm \nu}}}
\newcommand{\bmphi}{{\bm \phi}}
\newcommand{\tbmA}{\tilde{\bf {A}}}
\newcommand{\tbmB}{\tilde{\bf {B}}}

\newcommand{\bmmu}{{\bm \mu}}
\newcommand{\bmSigma}{{\bm \Sigma}}
\newcommand{\bmgamma}{{\bm \gamma}}

\newcommand{\bmlambda}{{\bm \lambda}}
\newcommand{\bmLambda}{{\bm \Lambda}}
\newcommand{\bmpsi}{{\bm \psi}}
\newcommand{\bmUpsilon}{{\bm \Upsilon}}
\newcommand{\bmOmega}{{\bm \Omega}}

\newcommand{\bfA}{{\bf A}}

\newcommand{\upd}{{\textup{d}}}
\usepackage{subfig}
\IEEEoverridecommandlockouts
\usepackage{algpseudocode}
\usepackage{graphicx,color}
\usepackage{epsfig}
\usepackage{amsmath,amssymb,amsfonts}
\usepackage{epstopdf}
\usepackage[linesnumbered,ruled,vlined]{algorithm2e}
\SetKwInput{KwInput}{Input}                
\SetKwInput{KwOutput}{Output}              
\SetKwProg{Initialize}{Initialize}{}{} 
\usepackage{mathtools}
\newcommand{\af}{{\textup {af}}}

\begin{document} 
\title{{Multipath Component Power Delay Profile Based Joint Range and Doppler Estimation for AFDM-ISAC Systems}
\thanks{Part of the content of this paper was previously presented at \cite{XiaoPara}.}

}

\author{%
	
	\IEEEauthorblockN{Fangqing Xiao, Zunqi Li, Dirk Slock}
	\IEEEauthorblockA{Communication Systems Department\\ 
		Eurecom, France\\
		Email: \{fangqing.xiao, zunqi.li, dirk.slock,\}@eurecom.fr}
}

\maketitle
\begin{abstract}
Integrated Sensing and Communication (ISAC) systems combine sensing and communication functionalities within a unified framework, enhancing spectral efficiency and reducing costs by utilizing shared hardware components. This paper investigates multipath component power delay profile (MPCPDP)-based joint range and Doppler estimation for Affine Frequency Division Multiplexing (AFDM)-ISAC systems. The path resolvability of the equivalent channel in the AFDM system allows the recognition of Line-of-Sight (LoS) and Non-Line-of-Sight (NLoS) paths within a single pilot symbol in fast time-varying channels. We develop a joint estimation model that leverages multipath Doppler shifts and delays information under the AFDM waveform. Utilizing the MPCPDP, we propose a novel ranging method that exploits the range-dependent magnitude of the MPCPDP across its delay spread by constructing a Nakagami-m statistical fading model for MPC channel fading and correlating the distribution parameters with propagation distance in AFDM systems. This method eliminates the need for additional time synchronization or extra hardware. We also transform the nonlinear Doppler estimation problem into a bilinear estimation problem using a First-order Taylor expansion. Moreover, we introduce the Expectation Maximization algorithm to estimate the hyperparameters and leverage the Expectation Consistent algorithm to cope with high-dimensional integration challenges. Extensive numerical simulations demonstrate the effectiveness of our MPCPDP-based joint range and Doppler estimation in ISAC systems.
\end{abstract}
\section{Introduction}

\begin{figure}
    \centering
    \includegraphics[width=1\linewidth]{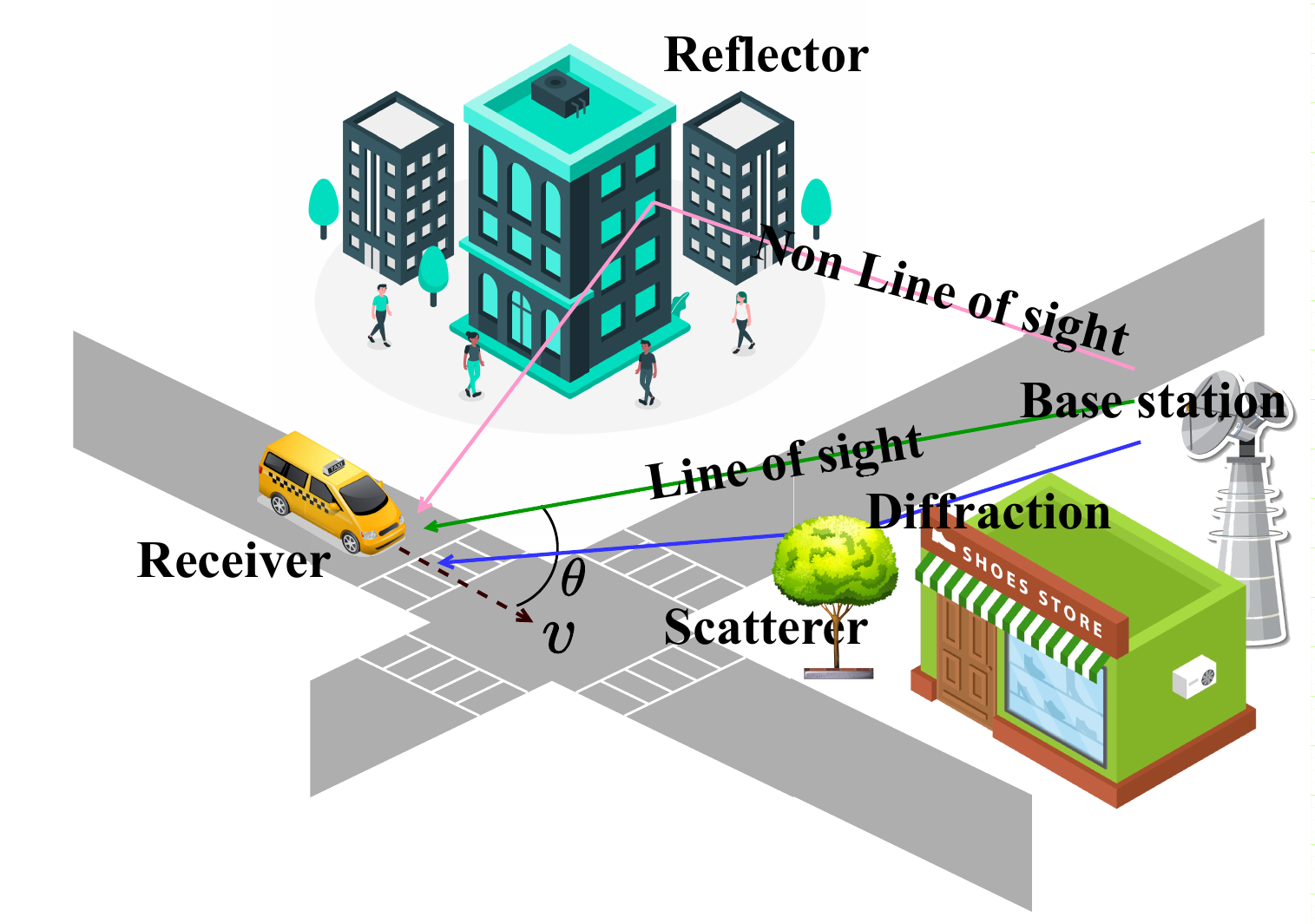}
    \caption{Example of LoS and NLoS links of a Moving Car}
    \label{fig:MPCEnv}
\end{figure}

\IEEEPARstart{I}{ntegrated} sensing and communication (ISAC) systems are transforming next-generation wireless networks by combining communication and sensing functionalities into a unified framework \cite{IntegratedLiu, IntegratingCui, IntegratedLu}. This integration improves spectrum efficiency and reduces costs by utilizing common components such as Radio Frequency (RF) front-ends and signal processing modules. ISAC systems play a crucial role in various applications including smart homes \cite{JointHuang}, industrial automation \cite{WirelessIslam}, and autonomous transportation \cite{IntegratedCheng}, where accurate sensing of distance, velocity, and direction is essential. Traditional sensing methods, such as LIDAR \cite{ALiDARDou}, RADAR \cite{WirelessLai}, and cameras \cite{InformationLiu}, face challenges like adverse weather, poor lighting, and high costs. In contrast, wireless signals from technologies such as WiFi and 5G New Radio (NR) \cite{Cha2024} provide effective alternatives for sensing. By harnessing these signals, ISAC systems meet the needs of future wireless networks, aligning sensing capabilities with communication advancements.

\subsection{Promising ISAC waveforms from OFDM to AFDM}
With advancements in vehicle-to-everything (V2X) and Unmanned Aerial Vehicle (UAV) networks, traditional Orthogonal Frequency Division Multiplexing (OFDM) waveforms struggle in dynamic environments due to severe inter-carrier interference. This challenge has spurred the development of innovative physical-layer waveforms, specifically orthogonal time frequency space (OTFS) \cite{L1} and affine frequency division multiplexing (AFDM) \cite{AffineAli}.

Both AFDM and OTFS spread signals across the complete time-frequency resource to mitigate selective fading \cite{L1}. As highlighted in \cite{AffineAli}, these techniques greatly improve transmission reliability by exploiting path diversity gains. Notably, AFDM offers a spectral efficiency advantage over OTFS, mainly due to its reduced need for pilot guard intervals \cite{AffineAli,L3}. Its flexible parameter settings further enable advanced encryption features, enhancing communication security \cite{L4}.

Moreover, AFDM's distinctive orthogonal chirp-based subcarrier structure not only promotes robust data transmission but also boosts sensing capabilities \cite{L5,L6}. This dual functionality makes AFDM a superior choice for ISAC waveforms \cite{L7}, delivering significant benefits for future wireless systems that demand both high-performance communication and accurate environmental sensing.

\subsection{MPCPDP-based Ranging Formulation}
Consequently, the interest in using wireless signals for joint determination of line-of-sight (LoS) distance and velocity relative to moving objects along the LoS path has surged, driven by the demand for high accuracy in sensing applications. However, as depicted in Fig. $\ref{fig:MPCEnv}$, complex environments often cause receivers to encounter signals arriving via multiple paths, both LoS and non-line-of-sight (NLoS), which exacerbates the multipath effect. To overcome these challenges, extensive research has been conducted on methods to either mitigate or exploit the Channel State Information (CSI) to enhance sensing \cite{dardari2009ranging, wymeersch2012machine, bialer2013maximum, kazaz2022delay, venus2023graph, venus2022graph, Li2022, leitinger2015evaluation, gentner2016multipath, leitinger2019belief, leitinger2019multipath, kim2022pmbm, PapakonstantinouGlobecom2008, PapakonstantinouPIMRC09}. For instance, Dardari et al. \cite{dardari2009ranging} introduced a maximum-likelihood estimator (MLE) for Time of Arrival (ToA)-based sensing, along with practical low-complexity schemes to reduce interference. Wymeersch et al. \cite{wymeersch2012machine} developed a machine learning technique to address bias in both LoS and NLoS conditions. Further, Li et al. \cite{Li2022} proposed a belief propagation (BP)-based algorithm for sequential channel estimation and detection (CEDA) of multipath component (MPC) parameters, like distance and angle of arrival (AoA), using radio signals. Building on these advancements, Venus et al. \cite{venus2023graph, venus2022graph} explored scenarios with obstructed LoS, aiming for precise position estimation. These methods, while effective, often rely on additional data such as ToA or AoA, requiring extra hardware or precise synchronization, thus increasing system complexity. An alternative, the use of received signal strength (RSS) for sensing \cite{AhmedSAS, yang2013RSSi, zanca2008experimental}, avoids the need for extra hardware. RSS methods, which measure the relationship between signal attenuation and distance, offer a simpler setup for ranging but are particularly susceptible to multipath interference in complex settings.

In our previous work \cite{MultipathXiao}, we proposed a novel approach that combines the benefits of RSS-based and CSI-based methods, employing the MPC Power Delay Profile (MPCPDP) for line-of-sight (LoS) distance estimation in orthogonal frequency-division multiplexing (OFDM) systems \cite{graff2021analysis}. We developed a statistical fading model of the PDP, establishing a relationship between the distribution parameters and propagation distance. Compared to traditional statistical models used in RSS-based methods—such as Rayleigh, Rician, or log-normal distributions—the Nakagami-m distribution \cite{stefanovic2013some} proved more versatile and accurate, fitting a broad range of experimental data \cite{abuelenin2018similarity}. This superiority arises from its ability to model the superposition of primary and diffuse reflected signals within a single path, thereby offering a better fit than the Rayleigh distribution \cite{stefanovic2011some}. Notably, near their mean values, Nakagami-m and Rician distributions exhibit similar behaviors. Consequently, we have validated the Nakagami-m decay model for our MPCPDP-based ranging method, which has shown enhanced resistance to multipath interference compared to conventional RSS methods. Although it lacks the precision of Time-of-Arrival (ToA)-based methods, it notably reduces the need for time synchronization or additional hardware, unlike other CSI-based techniques. Moving forward, a significant challenge remains: extending our approach to simultaneously estimate Doppler shifts in time-varying channel environments.

\subsection{Contributions}

Building on this foundation, we extend our MPCPDP-based ranging approach to include sensing capabilities under AFDM systems. This expansion enables the joint estimation of distance and velocity along the LoS propagation path, fulfilling crucial sensing requirements. By capitalizing on the attenuation characteristics captured in the MPCPDP, our proposed method eliminates the dependence on ToA or DoA information, allowing for effective sensing from a single base station. Additionally, this approach simplifies the implementation process while ensuring high accuracy. The key contributions of this paper are as follows:
\begin{itemize}
    \item \textbf{Adapting MPCPDP-based Ranging to AFDM for Enhanced Mobility Sensing}: Expanding on our previous research, this paper explores adapting the MPCPDP-based ranging technique to AFDM systems, aiming to improve sensing capabilities in mobile contexts. Assuming amplitude attenuation of each path adheres to a Nakagami-m distribution, our AFDM model accurately estimates both the LoS distance and Doppler shift. This approach notably cuts pilot overhead, thus reducing communication delays and markedly enhancing performance over traditional OFDM-based methods.

    \item \textbf{Developing an AFDM-based Joint Estimation Framework}: By utilizing the AFDM module, we have derived a mathematical model for the Doppler frequency shift and channel characteristics. Employing the Maximum-Likelihood Estimator (MLE), we outline the estimation procedures for both distance and Doppler shift. For distance estimation, the Expectation Maximization-Expectation Consistent (EM-EC) algorithm is applied to tackle the challenges posed by unknown hidden variables and complex high-dimensional integration, especially when the Nakagami-m distribution does not simplify to the Rayleigh distribution. For the Doppler shift estimation, we use a first-order Taylor expansion to refine the Doppler shift formula. This allows for the closed-form estimation of an object's moving speed relative to the LoS path, thereby enhancing our system’s perceptual capabilities.

       \item \textbf{Performance Metrics and Simulation Insights for MPCPDP Sensing}: When the Nakagami-m model simplifies to the Rayleigh model, the higher-dimensional integrals become computable, allowing us to derive Cramer-Rao Bounds (CRBs) for estimating LoS distance and velocity in the LoS propagation direction. These CRBs serve as a theoretical performance metric for our MPCPDP-based Sensing system. For the other cases,, deriving theoretical performance measures like CRBs becomes unfeasible. However, by analyzing the EM-EC algorithm, we identify the fixed point of MPCPDP-based Sensing. While this fixed point lacks a precise performance index, it aids further analysis. Additionally, we validate the feasibility of MPCPDP-based Sensing through simulations, which also establishes a solid foundation for future expansions of our work.\end{itemize}

The remainder of this paper is organized as follows: Section \ref{secII} introduces the system model, including the AFDM framework. Section  \ref{secIII} describes the MPCPDP-based sensing formulation. Section  \ref{secIV} presents the algorithm used for MPCPDP-based sensing. Section  \ref{secV} discusses the performance and fixed point analysis of the MPCPDP-based sensing method. Section  \ref{secVI} examines the simulation results, and Section  \ref{secVII} concludes the paper.

\subsubsection*{Notations}
Lowercase bold letters represent column vectors (e.g., $\bmx$), and uppercase bold letters represent matrices (e.g., $\bmX$). Scalars are in plain text (e.g., $x$). Individual entries of vectors and matrices are shown as $x_i$ and $X_{ij}$, respectively. The $\diag(\cdot)$ function either extracts diagonal elements of a matrix or forms a diagonal matrix from a vector. Probability densities are denoted by $p(\bmx)$. Operations for transpose, Hermitian transpose, and complex conjugation are $(\cdot)^T$, $(\cdot)^H$, and $(\cdot)^*$, respectively. Absolute value or modulus and vector norms are shown as $|\cdot|$ and $|\cdot|$, respectively. $\blkdiag(\bmA_1, \cdots, \bmA_N)$ constructs a block diagonal matrix. $\bmI_N$ and $\bmzero_N$ indicate an identity matrix and a zero vector of dimension $N$. $\kappa_{li}$ is the Kronecker delta, $\odot$ denotes the Hadamard product, and $j$ represents $\sqrt{-1}$.
\section{AFDM System Model} \label{secII}

\subsection{AFDM transmitter}
In the  AFDM  system transmitter \cite{AffineAli}, the baseband processing maps the input bit stream to an $M$-ary PSK/QAM constellation, resulting in a affine domain symbol sequence $\mathbf{x}$. Subsequently, the affine domain information is transformed into the time domain through an  Inverse Discrete Affine Fourier Transform (IDAFT) operation, expressed as:  

\begin{equation}
s_n = \frac{1}{\sqrt{N}} \sum_{m=0}^{N-1} e^{j 2\pi \left( c_1 n^2 + \frac{1}{N} m n + c_2 m^2 \right)} x_m, n=0,1,...,N-1 
\label{eq:IDAFT}
\end{equation}
where \( s_n \) represents the $N$-length block signal in time-domain, and \( e^{j 2\pi \left( c_1 n^2 + \frac{1}{N} m n + c_2 m^2 \right)} \) is the chirp orthogonal basis for the inverse discrete affine Fourier transform (IDAFT). Here, \( c_1 \) and \( c_2 \) are the chirp parameters that determine the time-frequency slope of the AFDM chirp subcarriers. 
Specifically, when \( c_1 = c_2 = 0 \), the time-frequency slope becomes zero, and AFDM degenerates  to the traditional multicarrier OFDM system.

To facilitate the representation, \eqref{eq:IDAFT} can be expressed in matrix form as:
\begin{equation}
\mathbf{s} = \bfA_{\af}^\mathrm{H} \mathbf{x}=({\bf{\Lambda} _{{c_2}}} \bf{F} {\bf{\Lambda} _{{c_1}}})^\mathrm{H} \mathbf{x},
\end{equation}
where $\mathbf{A}_{\af}^\mathrm{H}$ is the IDAFT matrix, $\bf{F}$ denotes the discrete Fourier matrix, and the  diagonal matrices ${\bf{\Lambda} _c}$ is defined as ${\bf{\Lambda} _c} = {\rm{diag}}({e^{-j2\pi c{n^2}}},  n = 0,1, \ldots, N - 1)$  .

\subsection{Channel model}
After adding a cyclic prefix (CP), the time-domain signal is transmitted through a time-varying channel. Without loss of generality, the transmitted signal propagates through the channel and is accompanied by additive white Gaussian noise (AWGN), resulting in the received signal:  
\begin{equation}
y(t) = \int h(t,\tau)s(t-\tau)d\tau + n(t),
\label{htt}
\end{equation}
where $h(t, \tau)$ denotes the complex impulse response of the channel. However, representing the channel in the delay-time domain fails to establish a direct connection with the physical propagation environment and does not simplify channel modeling for high-mobility scenarios.

To address these limitations, we focus on the delay-Doppler channel representation, which bridges the physical characteristics of the channel with electromagnetic wave propagation. In this model, a wireless channel is characterized by $L$ paths, including a line-of-sight (LoS) path, where each path is associated with distinct delay (caused by distance) and Doppler shift  (resulting from relative motion) parameters determined by the scatterers. The transmitted and received signals in (\ref{htt}) can thus be described using the delay-Doppler representation as:  
\begin{equation}
y(t) = \int \int h(\tau, \nu) \, s(t - \tau) e^{j 2 \pi \nu t} \, d\tau \, d\nu + n(t),
\end{equation}
where the delay-Doppler channel response $h(\tau, \nu)$ can be expressed as:  
\begin{equation}
h(\tau, \nu) = \sum_{i=1}^{L} h_i \delta(\tau - \tau_i) \delta(\nu - \nu_i), \label{eq:taui}
\end{equation}  
and $L$ represents the number of multipath components, $h_i$ is the complex path gain of the $i$-th path, $\tau_i$ denotes the delay of the $i$-th path, and $\nu_i$ represents the Doppler shift of the $i$-th path. For the LoS path, $i = 1$. 

According to \cite{AffineAli}, time domain channel matrix is given by
\begin{align}
{\bf{H}}_{\it{t}} = \sum\limits_{i = 1}^{L} {{h_i}} {{\bf{\Delta}} _{{\nu_i}}} {{\bf{\Pi}} ^{{l_i}}}, 
\label{eq:Hi}
\end{align}
where ${{\bf{\Delta}} _{{\nu_i}}} = {{\diag(0, e^{ - j\frac{{2\pi }}{N}{\nu_i}}},\cdots, e^{ - j\frac{{2\pi }}{N}{\nu_i}(N-1)}})$, $L$ represents the number of propagation paths, $\bf{\Pi}$ is the  cyclic shift matrix, and $l_i$ is the normalized delay tap of $i$-th path.

In wideband communication scenarios, the delay resolution can be considered sufficiently high. In this case, each propagation path corresponds to a distinct integer delay tap. The relative propagation distance $\Delta d_i$ and  the Doppler shift $\nu_i$ of $i$-th path can then be further described as:
\begin{align}
 {\Delta d_i} &= d_i-d_1= c \tau_i = c l_i /(N\Delta f), \\
\nu_i &= \frac{v \cos(\theta_i) f_c}{c \Delta_f} 
\end{align}   
where $f_c$ is the carrier frequency, $v$ is the velocity of the terminal,  $c$ is the speed of light, ${\theta_i}$ is the moving angle and uniformly distributed in the interval $[-\pi, \pi]$, and $\Delta f$ is the chirp subcarrier spacing.

 \subsection{{AFDM receiver}}
At the receiver, the AFDM system removes the CP from the received signal and applies an Discrete Affine Fourier Transform (AFT)—the inverse operation of (\ref{eq:IDAFT})—to convert the time-domain received signal back into the affine domain:
\begin{equation}
\bmy_{\af}=\bfA_{\af} \bmy= \bfA_{\af}(\bfH_t \bfA^\mathrm{H}_{\af} \bmx+ \bmn)=\bfH_{\af} \bmx+ \tilde{\bmn}  \label{eq:modelfrequence1}
\end{equation} 
Where the affine fourier domain equivalent channel $\mathbf{H}_{\af}= \bfA_{\af}\bfH_t \bmA_{\af}^\mathrm{H}$ and ${\tilde{\bmn} =\bfA_{\af} \bmn}$. Due to the Unitary of the AFT, $\tilde{\bmn}$ exhibits the same statistical properties as $\bmn$. Therefore, for the sake of simplicity, we do not change the special notation for them subsequently, and uniformly use $\bmn$ to represent the noise vectors in the time domain and affine frequency domain.

By substituting \eqref{eq:Hi} into \( \mathbf{H}_{\text{af}} \), \( \mathbf{H}_{\text{af}} \) can be decomposed into a sum of multipath components:
\begin{align}
\mathbf{H}_{\text{af}} &= \sum_{i=1}^{L} h_i\mathbf{H}_i =\sum_{i=1}^L h_i \mathbf{A}_{\af}   {{\bf{\Delta}} _{{k_i}}} {{\bf{\Pi}} ^{{l_i}}} \mathbf{A}_{\af}^\mathrm{H},  \label{eq:modelfrequence2}
\end{align}
where each element of \(\mathbf{H}_i\) can be specifically expressed as:

\begin{align}
\mathbf{H}_i (a, b)= \frac{1}{N} e^{\frac{j 2\pi}{N} \left( N c_1 l_p^2 - b l_p + N c_1 (a^2 - b^2) \right)} \mathbf{T}_i(a, b),
\end{align}
where:
\begin{align}
\mathbf{T}_i(a, b) = \sum_{n=0}^{N-1} e^{-\frac{j 2\pi}{N} \left[ (a-b + \nu_i + 2 N c_1 l_i)n \right]}\nonumber\\
 = \frac{e^{-i 2\pi (a-b + \nu_i  + 2 N c_1 l_i)} - 1}{e^{-\frac{j 2\pi}{N} (a-b + \nu_i + 2 N c_1 l_i)} - 1} .
\end{align}

\begin{figure}
    \centering
    \includegraphics[width=1\linewidth]{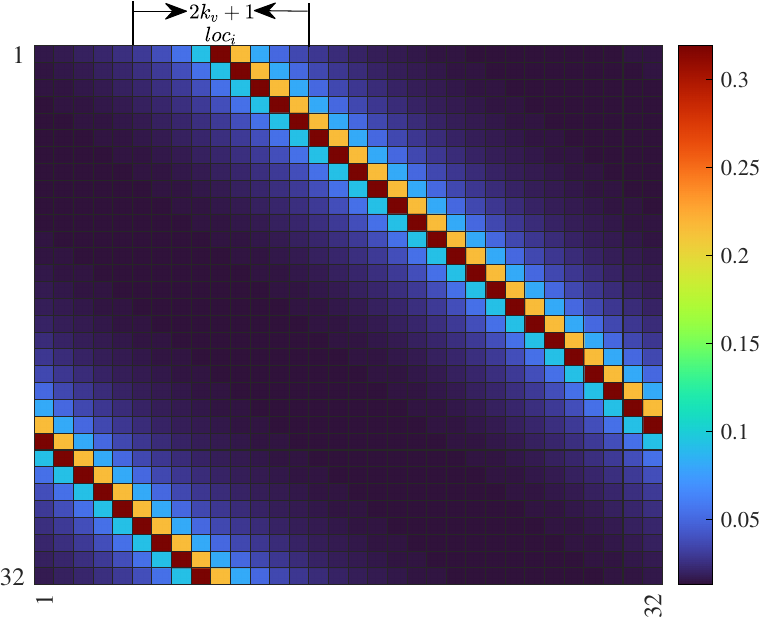}
    \caption{$\mathbf{H}_i$ under the scenario of fractional Doppler shift}
    \label{fig:Hp}
\end{figure}

Under the scenario of fractional Doppler shift, We can consider \( \mathbf{T}_i[a, b] \)  to be non-zero only within the interval centered at \( b = [a + \text{loc}_i]_N \), with a range of \( 2k_v + 1 \) values \cite{AffineAli} and $\text{loc}_i=[2 N c_1 l_i]_N$, \( k_v \) is  the sensitivity threshold, and the notation $[b] _N $ means taking the modulo operation of $b$ with respect to $N$. In the case of fractional Doppler shift, the matrix \( \mathbf{H}_i \) is illustrated in Fig. \ref{fig:Hp}.

 \subsection{Comparison of equivalent channel matrix between AFDM and OFDM}

From the above analysis, the positions of the non-zero elements in \( \mathbf{H}_i \) are determined by \(\text{loc}_i\), which depends on both the chirp parameter \( c_1 \) and the delay tap \( l_i \). Since AFDM allows adjustment of the non-zero band positions by setting a non-zero \( c_1 \), the equivalent matrix \( \mathbf{H}_{\text{af}} \) can distinguish multiple different paths. In contrast, OFDM is a special case of AFDM with \( c_1 = c_2 = 0 \), where \(\text{loc}_i=0 \) for any \( i \). Consequently, the non-zero positions of different paths in  OFDM system overlap near the diagonal, making them inseparable, as illustrated in Fig. 2. \begin{figure}[!t].
	\centering
	\subfloat{
		 \includegraphics[width=0.5\linewidth]{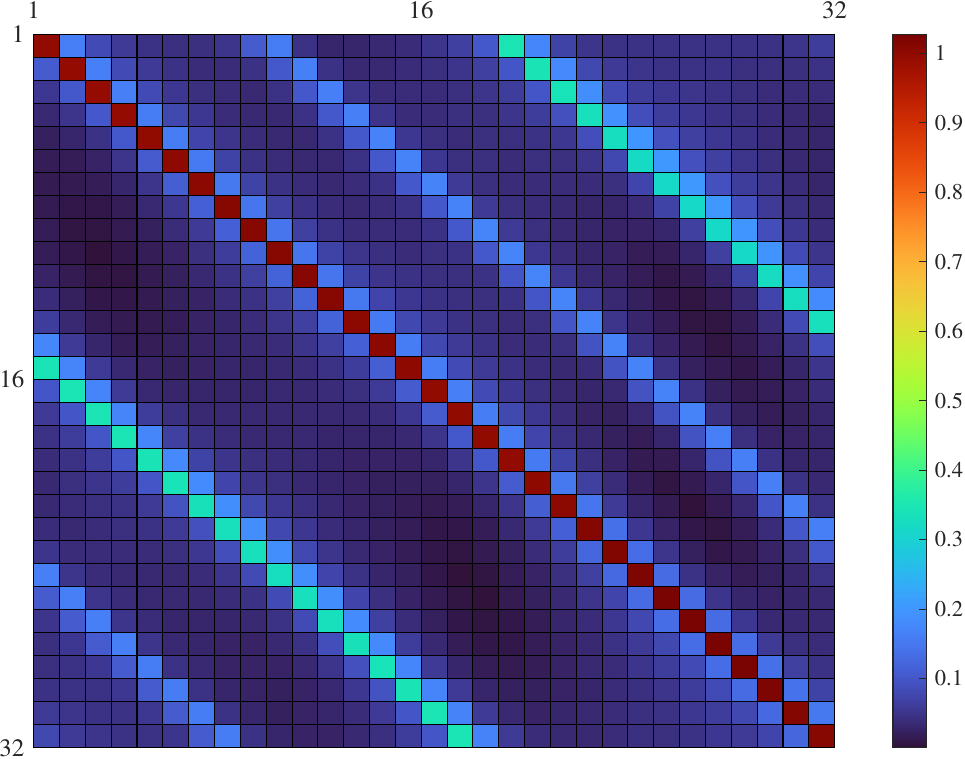}}
	\subfloat{
		\includegraphics[width=0.5\linewidth]{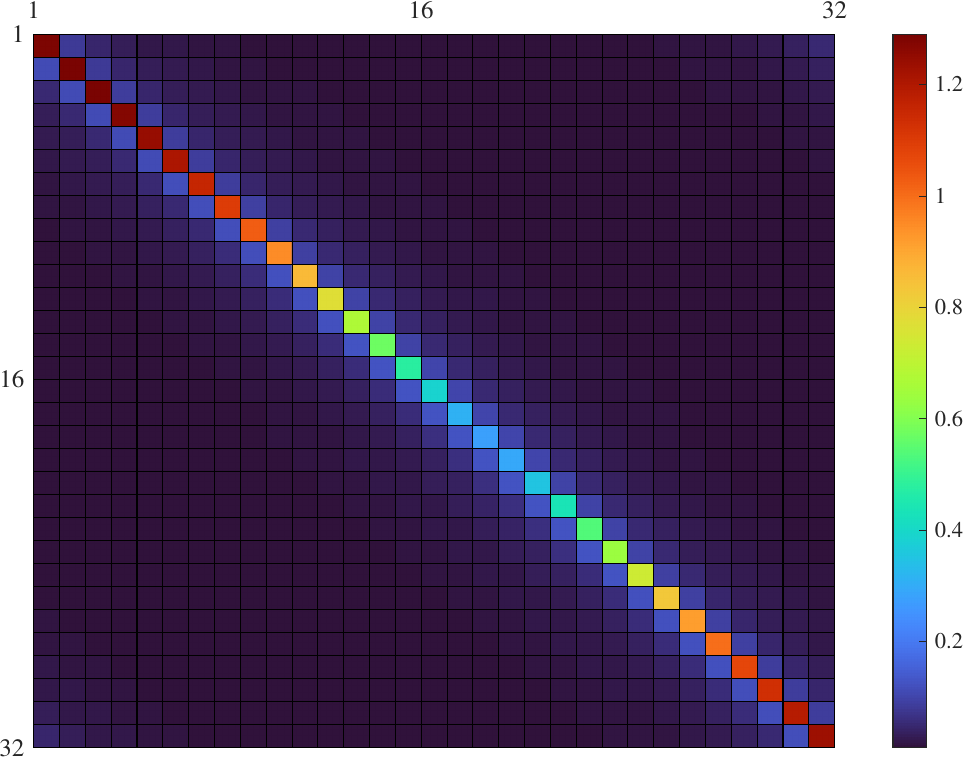}}
	\caption{Comparison of equivalent channel matrix between (a) AFDM and (b) OFDM ($N=32,L=3$).}
	\label{HafHf}
\end{figure} 

The path resolvability of the equivalent channel in AFDM demonstrates its capability to achieve fast path parameter estimation within a single symbol duration. In contrast, OFDM necessitates joint estimation across multiple consecutive symbols to resolve multipath components, resulting in higher sensing overhead and processing latency compared to AFDM. By strategically embedding pilot symbols and guard intervals, AFDM enables maximum likelihood (ML)-based estimation of path-specific parameters \cite{AffineAli}. Leveraging this intrinsic advantage, our subsequent theoretical analysis assumes perfect prior knowledge of both the number of multipath components and the relative delays between LoS  and NLoS paths. Consequently, this work does not prioritize parameter estimation algorithms for these quantities, as they are treated as deterministic inputs.
However, to rigorously validate system robustness, we later re-examine practical scenarios in simulations where the estimated number of paths is underestimated relative to the ground truth. 

\section{AFDM-based Sensing Framework} \label{secIII}
\subsection{Channel and Doppler Shift Estimation Model}
Based on the pre-estimated numberof paths $L$ and delays $\tau_i$ as described in \eqref{eq:taui}, and in conjunction with $\eqref{eq:modelfrequence2}$, \eqref{eq:modelfrequence1} can be rewritten as:
\begin{equation}
    \bmy_{\af} = {\bf{S} \bmR(\bmnu)} \bmh + \bmn, \label{eq:model1}
\end{equation}
where $\bmh = [h_1, \cdots, h_{L}]^T$, ${\bf{S}} =\mathbf{A}_{\text{af}}  [\diag({\bf{\Pi}}^0 \bms), \diag({\bf{\Pi}}^1 \bms), \cdots,  \diag({\bf{\Pi}}^{L-1} \bms)]$, ${{\bf{R}}(\bmnu)} = \blkdiag(\bmr(\nu_1), \bmr(\nu_2), \cdots,  \bmr(\nu_L))$, and $\bmr(\nu_l) = [1, e^{-\frac{j 2\pi \nu_i }{N}}, \cdots, e^{-\frac{j 2\pi \nu_i (N-1)}{N}}]^T$, $\bmn$ is in complex Gaussian distribution with zero mean and covariance $\sigma_n^2 \bmI_N$. 
Our objective is to estimate both $\bmnu$ and the hyperparameter $d_0$ of $\bmh$ jointly from \eqref{eq:model1} firstly and then estimate velocity $v$ and LoS angle $\theta_1$ from estimated $\bmnu$. 
We approximate the vector $\bmr(\nu_i)$ using first-order Taylor linearization:
\begin{equation}
	\bmr(\nu_i) \approx \bma(\tilde{\nu}_i) + \bmb(\tilde{\nu}_i) (\nu_i - \tilde{\nu}_i),
\end{equation}
where $\bma(\tilde{\nu}_i) = \bmr(\tilde{\nu}_i)$ and $\bmb(\tilde{\nu}_i) = \frac{\partial \bmr({\nu}_i)}{\partial \nu_i}\Big|_{\nu_i = \tilde{\nu}_i}$. By performing a first-order Taylor expansion with respect to (w.r.t.) ${\bf{R}(\bmnu)}$ around $\bmnu = \bmzero$, we obatin the following: 
\begin{equation}
        \bmy \approx \bmS (\bmA + \bmB \diag(\bmnu))\bmh + \bmn  = [\tbmA + \tbmB \diag(\bmnu)]\bmh + \bmn, \label{eq:updatetileAB}
\end{equation}
where $\bmA = \blkdiag[\bma(\tilde{\nu}_1 = 0), \cdots, \bma(\tilde{\nu}_L = 0)]$, ${\bmB} = \blkdiag[\bmb(\tilde{\nu}_1 = 0), \cdots, \bmb(\tilde{\nu}_L = 0)]$, $\tbmA = \bmS\bmA$ and $\tbmB = \bmS \bmB$. The variable $\bmh$ represents the complex attenuation coefficients, characterized by amplitude $|\bmh|$ and phase $\bmphi$ and the variable $\bmnu$ denotes the doppler shift. For each individual element $h_i = |h_i| e^{j\phi_i}$ of $\bmh$, the magnitude $|h_i|$ is modeled by a Nakagami-m distribution, while the phase $\phi_i$ follows a uniform distribution between $[-\pi, \pi]$. Consequently, the probability density functions (pdfs) of the magnitude and phase of $h_i$ are expressed as follows:
\begin{subequations}
    \begin{align}
        &p(|h_i| ; \Omega_i) = \frac{2 m^m |h_i|^{2m-1}}{\Gamma(m)\Omega_i^m} \exp\left[-\frac{m |h_i|^2}{\Omega_i}\right], m \geq 0.5;\\
        &p(\phi_i) = \frac{1}{2 \pi}, \phi_i \in [0, 2\pi),
    \end{align}
\end{subequations}
where $\Gamma(\cdot)$ denotes the gamma function, $\Omega_i$ is the average power intensity of the path $i$ and $m$ is the shape parameter of the Nakagami-m distribution. The shape parameter $m$ governs the fading characteristics of the distribution. For lower values of $m$, the distribution approximates a Rayleigh distribution with a faster decay, while higher values of $m$ result in a more concentrated distribution around its mean, indicating less severe fading. In practice, $m$ is typically estimated from channel measurements to accurately capture the fading characteristics of the specific wireless channel. According to the 3GPP model, the parameter $\Omega_i$ is defined as follows:
\begin{equation}
    \Omega_i(d_0) = P_t G_t G_r \left[ \frac{\lambda}{4\pi(d_0 + c \tau_i)}\right]^n_i = G_0 (d_0 + c\tau_i)^{-n_i}, \label{eq:Omega_id_0}
\end{equation}
in this given equation, several variables are defined as follows: \(P_t\) represents the transmitted power, \(G_t\) denotes the transmitting antenna gain, \(\lambda\) is the wavelength of the electromagnetic wave, \(c\) is the speed of light, \(n_i\) represents the propagation fading factor of the \(i\)-th path influenced by the environment, \(d_0\) indicates the LoS distance, and \(\tau_i\) denotes the propagation delay between the \(i\)-th path and the LoS path.

The term \(P_t G_t G_r \left[ \frac{\lambda}{4\pi} \right]^{n_i}\) can be treated as a constant, denoted as \(G_0\), which encapsulates the effects of transmit power, antenna gains, wavelength, and path loss exponent. The propagation fading factor \(n_i\) is critical in determining the rate of signal attenuation with distance, and it varies depending on the wireless channel characteristics and environmental conditions. As the propagation distance \(d_0 + c\tau_i\) increases, \(\Omega_i\) decreases according to an inverse power-law relationship, \((d_0 + c\tau_i)^{-n_i}\). This relationship enables the estimation of the specific range \(d_0\) based on \(\Omega_i\) when \(\tau_i\) is known for a given environment. 

\subsection{Estimate LoS range $d_0$ and related velocity $v \cos(\theta_1)$ in LoS direction }
Applying the jacobi determinant \cite{mirsky2012introduction}, the pdf of the complex fading coefficient $h_i$ can be derived as follows:
\begin{subequations}
    \begin{align}
        &p_{h_i}(h_i;\Omega(d_0, d_i)) = \frac{m^m |h_i|^{2m-2}}{\pi \Gamma(m) \Omega_i^m}\exp\left[-\frac{m |h_i|^2}{\Omega_i} \right]; \\
        &\Omega(d_0, d_i) = G_0 (d_0 + d_i)^{-n_i}, \label{eq:Omegad0}
    \end{align}    
\end{subequations}
For simplicity, we denote $p_{h_i}(h_i;\Omega(d_0, d_i))$ as $p_{a_i}(a_i; d_0)$. Accordingly, the pdf of the collection $\bmh$ is expressed as:
\begin{equation}
    p_{\bmh}(\bmh; d_0) = \prod_{i=1}^{L} p_{h_i} (h_i ; d_0). \label{eq:pbmhprior}
\end{equation}

The $\bmnu$ represents the Doppler shift caused by multipath propagation, where each component $\nu_i$ corresponds to the Doppler shift of the $i$-th path. Specifically, 
\begin{equation}
	v \cos(\theta_i) = \frac{\nu_i c \Delta_f}{f_c} , \label{eq:Dopplershift_i}
\end{equation}
where $v$ represents the target's absolute velocity, $\theta_i$ denotes the angle of movement along the $i$-th path. Our objective is to accurately estimate $d_0$ and $v \cos(\theta_1)$ based on estimated $\bmnu$, which corresponds to the velocity component in the LoS direction.
\subsection{MPCPDP-based Ranging and Doppler Shift Estimation}
Our objectives are twofold: first, to estimate $d_0$ and $\bmnu$ directly from $\bmy_{\af}$, and second, to derive $v\cos(\theta_1)$ from the estimated $\bmnu$. To achieve the first objective, we employ the joint maximum likelihood estimation (MLE) method, which reformulates the problem as follows:
\begin{equation}
    [\hat{d_0}, \hbmnu] = \arg \max_{d_0, \bmnu} \ell(d_0, \bmnu; \bmy_{\af}) = \arg \max_{d_0, \bmnu} \ln p(\bmy_{\af}; d_0, \bmnu), \label{eq:log-likelihood1}
\end{equation}
where $\ell(\cdot)$ denotes the log-likelihood function.
Regarding the optimization problem, the likelihood function in \eqref{eq:log-likelihood1} can be expressed as:
\begin{equation}
    p(\bmy_{\af}; d_0, \bmnu) = \int p_{\bmy_{\af}}(\bmy; \bmnu ) p_{\bmh}(\bmh; d_0) \upd \bmh. \label{eq:log-likelihood2}
\end{equation}
The pdf $p(\bmy_{\af}; d_0, \bmnu)$ is crucial for estimating the LoS range $d_0$ and $\bmnu$ based on received signal $\bmy_{\af}$. However, solving the integral problem directly to acquire $p(\bmy_{\af}; d_0, \bmnu)$ is intractable, as finding an analytical form poses significant challenges. Furthermore, the latent variable $\bmh$ is unobserved, and its distribution is unknown before reaching $d_0$. To tackle these challenges, the EM-EC algorithm is introduced later on. Moreover, The parameter $d_i$ representing the distance difference between the LoS path and the $i$-th path, also needs to be estimated. However, unless it happens to be an integer multiple of the sampling resolution, $d_i$ can never be accurately estimated.
\section{Expectation Maximization (EM) - Expectation Consistant (EC) Algorithm} \label{secIV}
\subsection{Review of Expectation Maximization}
Since direct optimization of \eqref{eq:log-likelihood1} is infeasible, we reformulate it into an iterative update process, laying the foundation for employing the Expectation-Maximization (EM) algorithm \cite{borman2004expectation, neal1998view}. Using the minorization-maximization (MM) framework \cite{hunter2004tutorial}, we construct a more tractable lower bound for the log-likelihood function in \eqref{eq:log-likelihood1}. 
Based on Bayes rules, we define the posterior pdf $p(\bmh| \bmy_{\af}, d_0, \bmnu)$ as follows:
\begin{equation}
	p(\bmh| \bmy_{\af}, d_0, \bmnu) = \frac{p_{\bmy_{\af}}(\bmy_{\af}; \bmh, \bmnu) p_{\bmh}(\bmh; d_0)}{\int p_{\bmy_{\af}}(\bmy_{\af}; \bmh, \bmnu) p_{\bmh}(\bmh; d_0) \upd \bmh}.
\end{equation}
At the $t$-th iteration, given the current estimates $d_0^{(t)}$, we can express the problem as follows:
\begin{subequations}
\begin{align}
    &\ell(d_0, \bmnu) - \ell(d_0^{(t)}, \bmnu^{(t)}) \\
    &= \ln \int p_{\bmy_{\af}}(\bmy_{\af}; \bmh, \bmnu) p_{\bmh}(\bmh; d_0) \upd \bmh - \ln p(\bmy; \bmnu^{(t)}, d_0^{(t)}) \\
    &= \ln \int \frac{p_{\bmy_{\af}}(\bmy_{\af}; \bmh, \bmnu) p_{\bmh}(\bmh; d_0)}{p(\bmh| \bmy_{\af}, d_0^{(t)}, \bmnu^{(t)})} p(\bmh| \bmy_{\af}, d_0^{(t)}, \bmnu^{(t)}) \upd \bmh \nonumber \\
    & - \int p(\bmh| \bmy_{\af}, d_0^{(t)}, \bmnu^{(t)}) \ln p(\bmy_{\af}; \bmnu^{(t)}, d_0^{(t)}) \upd \bmh \\
    & \geq \int p(\bmh| \bmy_{\af}, d_0^{(t)}, \bmnu^{(t)}) \nonumber\\
    & \quad \quad \quad \quad \quad \quad \times \ln \frac{p_{\bmy_{\af}}(\bmy_{\af}; \bmh, \bmnu) p_{\bmh}(\bmh; d_0)}{p(\bmy_{\af}; \bmnu^{(t)}, d_0^{(t)})  p(\bmh| \bmy_{\af}, d_0^{(t)}, \bmnu^{(t)})} \upd \bmh \\
    &= \E_{p(\bmh| \bmy, d_0^{(t)}, \bmnu^{(t)})} \left[  \ln \frac{p_{\bmy_{\af}}(\bmy_{\af}; \bmh, \bmnu) p_{\bmh}(\bmh; d_0)}{p(\bmy_{\af}; \bmnu^{(t)}, d_0^{(t)})  p(\bmh| \bmy_{\af}, d_0^{(t)}, \bmnu^{(t)})} \right]. \label{eq:EM1}
\end{align}
\end{subequations}
From \eqref{eq:EM1}, it follows that the updates for the update $d_0^{(t+1)}$ and $\bmnu^{(t+1)}$ can be derived as:
\begin{equation}
    [d_0^{(t+1)}, \bmnu^{(t+1)}] = \arg \max_{d_0, \bmnu} \E_{p(\bmh| \bmy_{\af}, d_0^{(t)}, \bmnu^{(t)})} \left[\ln p(\bmy_{\af}, \bmh; d_0, \bmnu) \right]. \label{eq:EM2}
\end{equation}
It is worth noting that the EM algorithm is guaranteed to converge to a (local) optimal point. Since $d_0$ and $\bmnu$ are embedded within $p_{\bmh}(\bmh; d_0)$ and $p_{\bmy_{\af}}(\bmy_{\af}; \bmh, \bmnu) $, respectively, a straightforward reorganization leads to the specific EM iteration expressed in \eqref{eq:EM2} as:
\begin{subequations}
    \begin{align}
        &d_0^{(t+1)} = \arg \max_{d_0} \E_{p(\bmh| \bmy_{\af}, d_0^{(t)}, \bmnu^{(t)})} [\ln p_{\bmh}(\bmh; d_0)]; \label{eq:optd0}\\
        &\bmnu^{(t+1)} = \arg\max_{\bmnu} \E_{p(\bmh| \bmy_{\af}, d_0^{(t)}, \bmnu^{(t)})} [\ln p_{\bmy_{\af}}(\bmy_{\af}; \bmh, \bmnu)]. \label{eq: optbmnu}
    \end{align}
\end{subequations}
For simplicity, we define the posterior mean and covariance of $\bmh$ under the pdf $p(\bmh| \bmy_{\af}, d_0^{(t)}, \bmnu^{(t)})$ as follows:
\begin{subequations}
    \begin{align}
        &\bmmu_{t} = \E_{p(\bmh| \bmy_{\af}, d_0^{(t)}, \bmnu^{(t)})}[\bmh]; \label{eq:posrealmean}\\
        &\bmSigma_{t} = \E_{p(\bmh| \bmy_{\af}, d_0^{(t)}, \bmnu^{(t)})}[(\bmh - \bmmu_{t})(\bmh - \bmmu_{t})^H]. \label{eq:posrealCovariance}
    \end{align} \label{eq:posreal}
\end{subequations}
It is worth noting that for the shape parameter $m \neq 1$ in the Nakagami-m distribution, solving \eqref{eq:posreal} becomes intractable. Therefore, the Expectation Consistent (EC) algorithm \cite{JMLRv6opper05a} is employed to address this challenge later on. Before proceeding, let us outline the optimization process for $d_0$ and $\bmnu$ given the approximated $\bmmu_t$ and $\bmSigma_t$.
\subsubsection{Optimize $d_0$}
Considering the pdf of $\bmh$, the EM iteration in \eqref{eq:optd0} can be reformulated as follows:
\begin{subequations}
    \begin{align}
        d_0^{(t+1)} &= \arg \min \sum_{i=1}^{L} \left[\ln \Omega_i(d_0) + \frac{\E_{p(\bmh| \bmy_{\af}, d_0^{(t)}, \bmnu^{(t)})} [|h_i|^2]}{\Omega_i(d_0, \hat{d}_i)} \right], \\
        &= \arg \min \sum_{i=1}^{L} \left[\ln \Omega_i(d_0) + \frac{(\bmmu_{t} \bmmu_{t}^H + \bmSigma_{t})}{\Omega_i(d_0, \hat{d}_i)} \right]
    \end{align} \label{eq:optd0formula}
\end{subequations}
In \eqref{eq:optd0formula}, $\Omega_i(d_0, \hat{d}_i)$ is defined in \eqref{eq:Omegad0}. Both $\ln \Omega_i(d_0, \hat{d}_i)$ and $\frac{1}{\Omega_i(d_0, \hat{d}_i)}$ are convex function w.r.t. $d_0$. This property ensures that the entire optimization function is convex, guaranteeing a unique global minimum point for $d_0$ at each time.
\subsubsection{Optimize $\bmnu$}
When considering the likelihood $p_{\bmy_{\af}}(\bmy_{\af}; \bmh, \bmnu)$, the EM iteration in \eqref{eq: optbmnu} can be transformed as follows:
\begin{subequations}
    \begin{align}
        &\bmnu^{(t+1)} = \arg \min_{\bmnu} \E_{p(\bmh| \bmy, d_0^{(t)}, \bmnu^{(t)})} \left[ \|\bmy_{\af} - [\tbmA + \tbmB \diag(\bmnu)]\bmh\|^2\right] \\
        &=\arg \min_{\bmnu}  \|\bmy_{\af} - [\tbmA + \tbmB \diag(\bmnu)]\bmmu_t\|^2 \nonumber\\
        &\quad \quad+ \tr \left\{ (\tbmA + \tbmB \diag(\bmnu)) \bmSigma_t (\tbmA + \tbmB \diag(\bmnu))^H \right\}. \label{eq:EMnu1}
    \end{align}
\end{subequations}
After straightforward algebraic manipulation, we obtain the first part in \eqref{eq:EMnu1}:
\begin{subequations}
    \begin{align}
        &\|\bmy_{\af} - [\tbmA + \tbmB \diag(\bmnu)]\bmmu_t\|^2 =\| (\bmy_{\af} - \tbmA \bmmu_t) - \tbmB \diag(\bmmu_t) \bmnu\|^2 \\
        &= \bmnu^T (\bmmu_t \bmmu_t^H \odot \tbmB^T \tbmB^*)\bmmu \nonumber\\
        &\quad \quad \quad   - 2\mathcal{R}\left\{\diag(\bmmu_t)\tbmB^H(\bmy_{\af} - \tbmA\bmmu_t) \right\}^T \bmnu + C_1,
    \end{align} \label{eq:EMnu2}
\end{subequations}
and the second part in \eqref{eq:EMnu1}:
\begin{subequations}
    \begin{align}
        &\tr\left\{ (\tbmA + \tbmB \diag(\bmnu)) \bmSigma_t (\tbmA + \tbmB \diag(\bmnu))^H \right\} \\
        &=2\mathcal{R}\left\{ \diag(\tbmB^H \tbmA \bmSigma_t)\right\}^T\bmnu + \bmnu^T(\bmSigma_t \odot \tbmB^T \tbmB^*) \bmnu + C_2,
    \end{align} \label{eq:EMnu3}
\end{subequations}
where $C_1$ and $C_2$ are constants independents of $\bmnu$.

For the sake of simplicity, we define $\bmP$ and $\bmgamma$ as follows:
\begin{subequations}
    \begin{align}
        &\bmP = \mathcal{R}\left\{ (\bmmu_t \bmmu_t^H + \bmSigma_t)  \odot \tbmB^T \tbmB^*\right\},\label{eq:EMP}\\
        &\bmgamma = \mathcal{R} \left\{ \diag(\bmmu_t^*)\tbmB^H(\bmy_{\af} - \tbmA\bmmu_t) - \diag(\tbmB^H\tbmA\bmSigma_t)\right\}.
    \end{align} \label{eq:EMnu4}
\end{subequations}
By combining \eqref{eq:EMnu2}, \eqref{eq:EMnu3}, and \eqref{eq:EMnu4} into \eqref{eq:EMnu1}, we can optimize $\bmnu$ as follows:
\begin{equation}
    \bmnu^{(t+1)} = \arg \min_{\bmnu} \bmnu^T \bmP \bmnu - 2\bmgamma^T \bmnu  = \bmP^{-1}\bmgamma, \label{eq:optbmnureal}
\end{equation}
Moreover, as $\bmP$ in \eqref{eq:EMP} is a positive semi-definite matrix, $\bmnu^{(t+1)} = \bmP^{-1}\bmgamma$ is a unique global optimal point.

\subsection{Expectation Consistent Approach}
As previously noted, the EM algorithm becomes intractable for a Nakagami-m prior (the shape parameter $m \neq 1$) due to the difficulty of obtaining the posterior mean and covariance in \eqref{eq:posreal}, which involves complex integration. Consequently, it is essential to develop an alternative algorithm that approximates the posterior distribution with a more tractable one. To address this challenge, we employ the expectation consistent (EC) algorithm.

Before describing the EC method, we introduce additional notation. Our goal is to approximate $p(\bmh|\bmy_{\af}, d_0, \bmnu)$ with $q(\bmh)$, which is chosen from an exponential family. The distribution  $q(\bmh)$ can be expressed as:
\begin{equation}
    q(\bmh; \bmlambda_q) = \frac{1}{Z_q} \exp(\bmlambda_q \bmg(\bmh)),
\end{equation}
where the partition functon $Z_q$ is obtained by integration as:
\begin{equation}
    Z_q = \int \exp (\bmlambda_q^T \bmg(\bmh)) d\bmh,
\end{equation}
With either initialized or optimized $[d_0, \bmnu]$, the EC algorithm attempts to calculate an estimated belief of the posterior pdf $p(\bmh; \bmy_{\af}, d_0, \bmnu)$ of the form of $r(\bmh)$ and $s(\bmh)$ as follows:
\begin{subequations}
    \begin{align}
        r(\bmh) = \frac{1}{Z_r} p_{\bmy_{\af}}(\bmy_{\af}|\bmh, \bmnu) \exp(\bmlambda_r^T \bmg(\bmh)),\\
        Z_r = \int p_{\bmy_{\af}}(\bmy_{\af}|\bmh, \bmnu) \exp(\bmlambda_r^T \bmg(\bmh)) d\bmh\\
        s(\bmh) = \frac{1}{Z_s} p_{\bmh}(\bmh; d_0) \exp(\bmlambda_s^T \bmg(\bmh))\\
        Z_s = \int p_{\bmh}(\bmh; d_0) \exp(\bmlambda_s^T \bmg(\bmh)) d \bmh
    \end{align}
\end{subequations}
where the function vector $\bmg(\bmh)$ is chosen to enable efficient and tractable computation of the required integrals ($Z_q$, $Z_r$ and $Z_s$), with the parameters $\bmlambda$ adjusted to optimize specific criteria. In this context, the terms "efficient" and "tractable" refer to a specific set of approximating functions $\bmg(\bmh)$. Typically, the i.i.d. complex Gaussian component remains effective and computationally feasible as long as $\bmg(\bmh)$ includes the first and second moments of $\bmh$. Furthermore, optimizing $d_0$ \eqref{eq:optd0formula} and $\bmnu$ in \eqref{eq:optbmnureal} require only the posterior first-order and second-order moments, reinforcing the suitability of the Gaussian assumption. Under this framework, $\bmlambda$ and $\bmg(\bmh)$ can be expressed as:
\begin{subequations}
    \begin{align}
        &\bmg(\bmh) = (2h_1, \cdots, 2h_L, -|h_1|^2, \cdots, -|h_L|^2)^T, \\
        &\bmlambda = (\eta_1, \cdots, \eta_L, \Lambda_1, \cdots, \Lambda_L)^T.
    \end{align}
\end{subequations}
\begin{algorithm}[t]
	\DontPrintSemicolon
	\KwInput{$\tbmA$, $\tbmB$, $\bmy_{\af}$, $\bmg(\bmlambda)$, $d_0$ and $\bmnu$}
	\KwOutput{$\bmlambda_q$}
	{Initialize: $\bmlambda_r$, $\bmlambda_q$, $\bmlambda_s$}\\
	\While{stopping criterion not fulfilled}
	{
		{// Sending message from r to s} \\
		{Solve $\bmlambda_q$ by $\E_{q}[\bmg(\bmh)| \bmlambda_q] = \E_{r}[\bmg(\bmh) |\bmy_{\af}, \bmlambda_r, \bmnu]$};\label{eq:bmlambdaq1}\\
		{$\bmlambda_s = \bmlambda_q - \bmlambda_r$}\\
		{// Sending message from s to r}\\
		{Solve $\bmlambda_q$ by $\E_{q}[\bmg(\bmh)| \bmlambda_q] = \E_{s}[\bmg(\bmh) |\bmlambda_s, d_0]$} \label{eq:bmlambdaq2}\\
		{$\bmlambda_r = \bmlambda_q - \bmlambda_s$} \\
	}
	\caption{Expectation Consistent Algorithm} \label{alg:EC}
\end{algorithm}
The detailed steps of the EC algorithm are outlined in Algorithm~\ref{alg:EC}. In lines \ref{eq:bmlambdaq1} and \ref{eq:bmlambdaq2}, these steps are commonly referred to as moment matching between $q(\bmh)$ and $s(\bmh)$ and $r(\bmh)$, respectively, as detailed below:
\begin{subequations}
\begin{align}
	&\E_{r}[\bmg(\bmh) |\bmy_{\af}, \bmlambda_r, \bmnu] = \frac{\int \bmg(\bmh)  p(\bmy|\bmh, \bmnu)\exp(\bmlambda_r^T \bmg(\bmh)) d \bmh }{\int p(\bmy_{\af}|\bmh, \bmnu)\exp(\bmlambda_r^T \bmg(\bmh)) d \bmh}, \label{eq:expectationr}\\
	&\E_{s}[\bmg(\bmh) |\bmlambda_s, d_0] = \frac{\int \bmg(\bmh)  p(\bmh|d_0)\exp(\bmlambda_s^T \bmg(\bmh)) d \bmh }{\int p(\bmh|d_0)\exp(\bmlambda_s^T \bmg(\bmh)) d \bmh}.\label{eq:expectations}
\end{align}
\end{subequations}
In \eqref{eq:expectationr}, since $p_{\bmy}(\bmy|\bmh, \bmnu)$ follows a complex Gaussian distribution, we can express it as:
\begin{subequations}
    \begin{align}
         &\E_{r}[\bmh] = (\diag(\bmLambda_r) + \sigma_n^{-2} \bm{\Delta}^H \bm{\Delta})^{-1} (\sigma_n^{-2} \bm{\Delta}^H \bmy + {\bmeta_r}); \\
         &\E_{r}[\bmh \bmh^H] = \E_{r}[\bmh]  \E_{r}[\bmh]^H + (\diag(\bmLambda_r) + \sigma_n^{-2} \bm{\Delta}^H \bm{\Delta})^{-1},
    \end{align} \label{eq:updateErhh}
\end{subequations}
where 
\begin{equation}
    \bm{\Delta} = \tbmA + \tbmB\diag{(\bmnu)};
\end{equation}
therefore, the $\bmLambda_q = [\bmeta_q^T, \bmLambda_q^T]^T$ in line \ref{eq:bmlambdaq1} can be calculated as:
\begin{subequations}
    \begin{align}
        &\bmLambda^q = (\E_{r}[\bmh \bmh^H] - \E_{r}[\bmh] \E_{r}[\bmh]^H)^{-1};\\
        &\bmeta^q = \bmLambda^q \odot \E_{r}[\bmh].
    \end{align} \label{eq:updateEqhh_r}
\end{subequations}
Moreover, in \eqref{eq:expectations}, as $p_{\bmh}(\bmh; d_0)$ is given in \eqref{eq:pbmhprior}, we can have
\begin{subequations}
    \begin{align}
        &\E_s[h_i] = \frac{m \Omega_i \eta_{s,i}}{m + \Lambda_{s,i}\Omega_i} \frac{_1F_1(m+1;2;\frac{\Omega_i |\eta_{s,i}^2|}{m + \Lambda_{s,i}\Omega_i})}{_1F_1(m;1;\frac{\Omega_i |\eta_{s,i}^2|}{m + \Lambda_{s,i}\Omega_i})}, \\
        &\E_s[|h_i|^2] = \frac{m \Omega_i \eta_{s,i}}{m + \Lambda_{s,i}\Omega_i} \frac{_1F_1(m+1;1;\frac{\Omega_i |\eta_{s,i}^2|}{m + \Lambda_{s,i}\Omega_i})}{_1F_1(m;1;\frac{\Omega_i |\eta_{s,i}^2|}{m + \Lambda_{s,i}\Omega_i})}, 
    \end{align} \label{eq:updateEshh}
\end{subequations}
where $_1F_1(a;b;z)$ represents the confluent hypergeometric
function, defined by the hypergeometric series:
\begin{equation}
    _1 F_1(a;b;z) = \sum_{k=0}^{+\infty} \frac{(a)_k z^k}{(b)_k k!};
\end{equation}
where
\begin{equation}
    (a)_0 = 1, (a)_k = a(a+1)(a+2)\cdots(a+k-1),
\end{equation}
 is the rising factorial. 
Similar to the approach used previously, the $\bmLambda_q = [\bmeta_q^T, \bmLambda_q^T]^T$ in line \ref{eq:bmlambdaq2} can be computed as:
\begin{subequations}
    \begin{align}
        &\bmLambda^q = (\E_{s}[\bmh \bmh^H] - \E_{s}[\bmh] \E_{s}[\bmh]^H)^{-1};\\
        &\bmeta^q = \bmLambda^q \odot \E_{s}[\bmh].
    \end{align} \label{eq:updateEqhh_s}
\end{subequations}
In addition, it can also be represented as the solution of minimum KL-divergence as bellow:
\begin{align}
	&q_r(\bmh)  = \mathop{\arg\min}_{q(\bmh)} D_{KL} [r(\bmh; \bmnu) \|q(\bmh)], \\
	&q_s(\bmh)  = \mathop{\arg\min}_{q(\bmh)} D_{KL} [s(\bmh; d_0) \|q(\bmh)].
\end{align}
The fixed point utilized in the EC algorithm can be expressed as:
\begin{equation}
	\E_r[\bmg(\bmh); \bmnu] = \E_s[\bmg(\bmh); d_0] =  \E_q[\bmg(\bmh)]. \label{eq: fixed point of EC}
\end{equation}
\subsection{MPCPDP-based Sensing Algorithm}
The algorithm for MPCPDP-based sensing is outlined in Algorithm \ref{alg:MPCPDP-based sensing}. Initially, vector $\bmnu$ is set to zero, and the $\bmOmega$ is set based on $d_0$ is set to be $10$ meters. The vectors $\bmlambda_r$ and $\bmlambda_q$ are initialized with all one. Instead of requiring the EC step to undergo a fixed number of iterations, the process is halted when the normalized difference $\|\bmlambda_q^{new} - \bmlambda_q^{new}\|^2/ \|\bmlambda_q^{new}\|$ falls below a tolerance threshold $\epsilon_1$. Similarly, in the EM algorithm, iterations cease when the normalized difference $\|d_0^{new} - d_0 \|^2/\|d_0^{new}\|^2$ is less than another tolerance threshold \(\epsilon_2\). 

\begin{algorithm}[t]
	\DontPrintSemicolon
	\KwInput{$\bmy$, $P_n$, $m$,  $n_i$, $\sigma_n^2$, $[\hat{\tau}_1, \cdots, \hat{\tau}_L]$}
	\KwOutput{$d_0$, $\bmnu$}
	{Initialize: $\bmnu$, $\bmOmega$}\\
	{Set $\tbmA$ and $\tbmB$ based on $\bmnu$ based on \eqref{eq:updatetileAB}}\\
	\While{EM stopping criterion not fulfilled}
	{
		{ // Applying EC algorithm to get $\bmOmega$} \\
		{Initialize: $\bmlambda_r$, $\bmlambda_s$}\\
		\While{EC stopping criterion not fulfilled}{	
			{Solve $\bmlambda_q$ based on \eqref{eq:updateEqhh_r}} \\
			{Update $\bmlambda_r$ based on \eqref{eq:updateErhh}}\\
			{Solve $\bmlambda_q$ based on \eqref{eq:updateEqhh_s}} \\
			{Update $\bmlambda_s$ based on \eqref{eq:updateEshh}}\\}
		{Update $\bmOmega$ based on $\bmlambda_q$} \\
		{ // Applying EM algorithm to get $d_0$ and $\bmnu$} \\
		{Optimize $d_0$ by $\bmOmega$ based on \eqref{eq:optd0formula}} \\
		{Optimize $\bmnu$ based on \eqref{eq:optbmnureal}} \\
	}
	\caption{MPCPDP-based Sensing Algorithm} \label{alg:MPCPDP-based sensing}
\end{algorithm}

\section{Theoretical Performance for the MPCPDP-based Sensing Method} \label{secV}
In this section, we explore the theoretical performance and fixed points of MPCPDP-based sensing. For cases where the shape parameter $m = 1$, we calculate the Cramer-Rao Bound (CRB) for the estimated parameters. Conversely, when $ m \neq 1 $, we identify the fixed point of the EM-EC algorithm, as the CRB becomes intractable due to the complexity of high-dimensional integration.
\subsection{Theoretical Cramer-Rao Bound (CRB) With the Shape Parameter $m = 1$ }
When \( m = 1 \), the path complex attenuation coefficients \( \bmh \) are independent and identically distributed (i.i.d.) complex Gaussian random variables with zero mean. Defining the estimated parameter as \( \bmpsi = [d_0, \bmnu^T]^T \), our objective is to estimate \( \bmpsi \) directly and solely from \( \bmy_{\af} \) using the maximum likelihood estimator based on the probability density function (pdf) of \( \bmy_{\af} \) given \( \bmpsi \). Through straightforward mathematical derivations, we obtain the log-likelihood function \( \ln p(\bmy; \bmpsi) \) as follows:
\begin{equation}
    \ln p(\bmy_{\af};\bmpsi) \propto -\ln \det (\bmUpsilon(\bmpsi)) - \bmy^H \bmUpsilon^{-1}(\bmpsi)\bmy,
\end{equation}
where
\begin{equation}
    \bmUpsilon(\bmpsi) = {\bf S \bmR(\bmnu)} \diag(\bmOmega(d_0))  {\bf\bmR}^H(\bmnu) {\bf{S}}^H  + \sigma_n^2 \bmI_N.
\end{equation}
Based on \cite{kay1993fundamentals}, the Fisher Information Matrix (FIM) of estimating $\bmpsi$ can be expressed as:
\begin{equation}
    \FIM(\bmpsi) = \E_0 \left[- \frac{\partial^2 \ln p_{pseudo}(\bmy;\bmpsi)}{\partial \bmpsi \partial \bmpsi^T} \right];
 \end{equation}
where $\E_0$ is taking expectation w.r.t. $p(\bmy_{\af}; \bmpsi)$. With some simple mathematical calculations, the FIM can be denoted as follows:
\begin{equation}
[\FIM(\bmpsi)]_{ij} = 
- \tr\left( \bmUpsilon^{-1} \frac{\partial \bmUpsilon}{\partial \psi_j} \bmUpsilon^{-1} \frac{\partial \bmUpsilon}{\partial \psi_i}\right), \label{eq:FIM1}
\end{equation} 
where
\begin{subequations}
    \begin{align}
    	&\frac{\partial \bmUpsilon}{\partial d_0} =  {\bf{S} \bmR(\bmnu)} \diag \left(\frac{ \partial {\bmOmega}(d_0)}{\partial d_0}\right)  {\bf \bmR}^H(\bmnu) {\bf{S}}^H \\
    	&\frac{\partial {\Omega}_i(d_0)}{\partial d_0} = -n_i G_0(d_0 + d_i)^{-n_i - 1};\\
    	&\frac{\partial \bmUpsilon}{\partial \nu_i} = {\bf{S}}\frac{\partial {\bf{R}}(\bmnu)}{\partial \nu_i}  \diag(\bmOmega(d_0)) {\bf\bmR}^H(\bmnu) {\bf{S}}^H \\
    	&\quad\quad\quad\quad\quad\quad+  {\bf S \bmR(\bmnu)}  \diag(\bmOmega(d_0))  \left(\frac{\partial {\bf{R}}(\bmnu)}{\partial \nu_i}\right)^H{\bf{S}}^H\\
    	&\frac{\partial {\bf{R}}(\bmnu)}{\partial \nu_i} =  [\blkdiag(\frac{\partial \bmr(\nu_1)}{\partial \nu_i}, \frac{\partial \bmr(\nu_2)}{\partial \nu_i}, \cdots,  \frac{\partial \bmr(\nu_L)}{\partial \nu_i})],\\
    	& \frac{\partial \bmr(\nu_l)}{\partial \nu_i}= \frac{-j 2 \pi \kappa_{li}}{N} [0, e^{-\frac{j 2\pi \nu_i }{N}}, \cdots, (N-1)e^{-\frac{j 2\pi \nu_i (N-1)}{N}}]^T,
    \end{align} \label{eq:FIM2}
\end{subequations}
In conclusion, using \eqref{eq:FIM1} and \eqref{eq:FIM2}, the CRB of $\bmpsi$ w.r.t. MPCPDP-based sensing can be calculated as follows:
\begin{equation}
	\CRB(\bmpsi) \triangleq \FIM^{-1}(\bmpsi).
\end{equation}
Consequently, we can derive the theoretical performance of the CRB jointly estimating the LoS range $d_0$ and the Doppler shift component $\nu_1$ as follows: 
\begin{subequations}
	\begin{align}
		&\CRB(d_0) = \CRB(\bmpsi)_{1,1};\\
		&\CRB(\nu_1) = \CRB(\bmpsi)_{2,2};
	\end{align}
\end{subequations}
Unfortunately, if the shape parameter $m \neq 1$, the high-dimensional integration required to be obtain the likelihood $p(\bmy_{\af};\bmpsi)$ is intractable, making it intractable to calculate its CRB. To solve the intractable high-dimension integration problem in EM, we introduce the EC algorithm, and we next analyze what the fixed point of the EM-EC algorithm would be.
\subsection{Fixed Point of the EM-EC Algorithm}
We now demonstrate that the parameter updates in the EM-EC algorithm for the MPCPDP-based sensing method can be interpreted as an approximation of the EM algorithm \cite{neal1998view}. The optimization function for EM-EC \cite{XiaoPara} is defined as:
\begin{equation}
	F(q, r, s, \bmtheta) \triangleq -D_{KL} \lsb r \| p(\bmy_{\af} | \bmh, \bmnu)\rsb -D_{KL} \lsb s \| p(\bmh; d_0)\rsb - H(q),
\end{equation}
where $\bmtheta = [d_0, \bmnu^T]^T$. It's also worth noting that $-F$ is commonly referred to as the energy equation in the EM-EC algorithm. However, by incorporating the constraint conditions derived from \eqref{eq: fixed point of EC}, commonly known as moment matching constraints, the optimization problem can be formulated as follows:
\begin{subequations}
\begin{align}
	\hat{\bmtheta} = \mathop{\arg \max}_{\bmtheta} \max_{r,s} \min_{q} F(q, r, s, \bmtheta)  \\
	{\rm{such\,to}}\quad \E_r[\bmg(\bmh); \bmnu]  =  \E_q[\bmg(\bmh)] \nonumber\\ \E_s[\bmg(\bmh); d_0]  =  \E_q[\bmg(\bmh)]. 
\end{align} \label{eq: optimization EM-EC}
\end{subequations}
It is important to note that the fixed points of the EM-EC algorithm align with the stationary points of the optimization problem defined in \eqref{eq: optimization EM-EC}. The Lagrangian for this constrained optimization in \eqref{eq: optimization EM-EC} is given by:
\begin{align}
	L(\bmtheta, q, r, s, \bmlambda_1, \bmlambda_2) \triangleq F(q, r, s, \bmtheta) + \bmlambda_1 (\E_r[\bmg(\bmh)| \bmnu]  \nonumber\\
	-  \E_q[\bmg(\bmh)]) + \bmlambda_2 (\E_s[\bmg(\bmh); d_0]  -  \E_q[\bmg(\bmh)]). \label{eq: EM-EC Larg}
\end{align}
In order, we first solve for $q(\bmh;\bmlambda_q)$ in \eqref{eq: EM-EC Larg} as
\begin{subequations}
\begin{align}
	&\hat{\bmlambda}_q = \mathop{\arg\min}_{{\bmlambda}_q}  L(\bmtheta, \hat{\bmlambda}_s, \hat{\bmlambda}_r, {\bmlambda}_q, \bmlambda_1, \bmlambda_2) \nonumber\\
	&= \mathop{\arg\min}_{{\bmlambda}_q} [\bmlambda_q^T - (\bmlambda_1 + \bmlambda_2)] \E_q[\bmg(\bmh)|\bmlambda_q].
\end{align}  \label{eq:EM-EC lambda_q}
\end{subequations}
By performing the first-order and second-order derivatives with respect to $\bmlambda_q$, we obtain:
\begin{align}
	&\frac{\partial L(\bmtheta, \hat{\bmlambda}_s, \hat{\bmlambda}_r, {\bmlambda}_q, \bmlambda_1, \bmlambda_2)}{\partial  \bmlambda_q} \nonumber\\
	&= [\bmlambda_q^T \! -\!(\bmlambda_1 + \bmlambda_2)^T] \left\{\E_q[\bmg(\bmh)\bmg(\bmh)^H] - \E_q[\bmg(\bmh)] \E_q[\bmg(\bmh)]^H\right\}, \\
	&\frac{\partial^2 L(\bmtheta, \hat{\bmlambda}_s, \hat{\bmlambda}_r, {\bmlambda}_q, \bmlambda_1, \bmlambda_2)}{\partial  \bmlambda_q \partial  \bmlambda_q^T}\nonumber \\
	&= \left\{\E_q[\bmg(\bmh)\bmg(\bmh)^H] - \E_q[\bmg(\bmh)] \E_q[\bmg(\bmh)]^H\right\}^T \geq 0,
\end{align}
therefore \eqref{eq:EM-EC lambda_q} is a convex function with only one minimum point $\hat{\bmlambda}_q$ as:
\begin{equation}
	\hat{\bmlambda}_q = \bmlambda_1 + \bmlambda_2. \label{eq: solution of q}
\end{equation} 
Next, we turn to solving for $s(\bmh; \bmlambda_s)$ and $r(\bmh; \bmlambda_r)$,
\begin{subequations}
\begin{align}
	&[\hat{\bmlambda}_s, \hat{\bmlambda}_r] = \mathop{\arg\max}_{{\bmlambda}_s, {\bmlambda}_r} L(\bmtheta, \bmlambda_s, {\bmlambda}_r, \hat{\bmlambda}_q, \bmlambda_1, \bmlambda_2) \\
	&= \mathop{\arg\max}_{{\bmlambda}_s, {\bmlambda}_r} (\bmlambda_1^T - \bmlambda_r^T ) \E_r[\bmg(\bmh)|\bmlambda_r] + (\bmlambda_2^T - \bmlambda_s^T ) \E_s[\bmg(\bmh)|\bmlambda_s].
\end{align} \label{eq:EM-EC sr}
\end{subequations}
Through simple algebraic analysis w.r.t. \eqref{eq:EM-EC sr}, it becomes evident that this function is concave and possesses fixed points as follows:
\begin{equation}
	\hat{\bmlambda}_r = \bmlambda_1, \quad \hat{\bmlambda}_s = \bmlambda_2.
\end{equation}
Finally, $\bmtheta = [d_0, \bmnu^T]^T$ can be optimized as: 
\begin{align}
	&\hat{\bmtheta} = [\hat{d}_0, \hat{\bmnu}] = \mathop{\arg\max}_{\bmtheta}  L(\bmtheta, \hat{\bmlambda}_s, \hat{\bmlambda}_r, \hat{\bmlambda}_q,  \bmlambda_1, \bmlambda_2) \nonumber\\
	&= \mathop{\arg\max}_{\bmnu} \E_r[\ln p(\bmy_{\af}|\bmh, \bmnu)] + \mathop{\arg\max}_{d_0} \E_s[\ln p(\bmh; d_0)].
\end{align}
We then have the following theorem:

\!\!\!\!\!\textbf{Theorem 1:} \textit{At any fixed points of the EM-EC algorithm, we have}:
\begin{align}
	&\bmlambda_1 = \hat{\bmlambda}_r, \quad \bmlambda_2 = \hat{\bmlambda}_s, \quad \hat{\bmlambda}_q = \bmlambda_1 + \bmlambda_2;\\
	&\hat{q}(\bmh) = \frac{\exp(\hat{\bmlambda}_q^T \bmg(\bmh))}{\int \exp(\hat{\bmlambda}_q^T \bmg(\bmh)) d \bmh};\\
	&\hat{r}(\bmh) = \frac{p(\bmy_{\af} | \bmh, \hat{\bmnu})\exp(\hat{\bmlambda}_r^T \bmg(\bmh))}{\int p(\bmy_{\af} | \bmh, \hat{\bmnu}) \exp(\hat{\bmlambda}_r^T \bmg(\bmh) )d \bmh}; \\
	&\hat{s}(\bmh) = \frac{p(\bmh | \hat{d_0})\exp(\hat{\bmlambda}_s^T \bmg(\bmh))}{\int p(\bmh | \hat{d}_0)\exp(\hat{\bmlambda}_s^T \bmg(\bmh)) d \bmh},
\end{align}
\textit{where $\hat{q}$, $\hat{r}$, and $\hat{s}$ denote the critical points of the Lagrangian in \eqref{eq: EM-EC Larg} that satisfy the moment matching constraints in \eqref{eq: fixed point of EC}. If the EM-EC algorithm converges, its limit points correspond to the local optima of the EM-EC auxiliary function.}

\section{Simulation Results} \label{secVI}
	\begin{table}[t]
        		\centering
        		\caption{Parameters setting}
        		\begin{tabular}{cc}
        			\hline
        			\hline
        			Parameter &  Value \\
        			\hline 
        			$G_0$ & 1.  \\
        $N$ & 512 \\
        $f_c$ & 90 GHz \\
        $\Delta f$& 15 kHz\\ 
        $L$& 3, 7, 11  \\
        $v$ & Ranging from $30$ km/h to $90$ km/h \\
        			SNR (dB) & Range from 0 to 30\\
        			$n_{1}$ (LoS) & 2.19 \\
       			$n_{i \neq 1}$ (NLoS) & 3.19 \\
        			$m$ & Ranging from 1 to 10 \\
        			$d_0$ (m)& 100. \\
        			$l_i$ (m) & Random between 1 to 20 \\
        			$\theta_i$ & Random between 0 to 2$\pi$\\
        			\hline
        			\hline
        		\end{tabular}
        		\label{tab:my_label}
        	\end{table}
In this section, we present simulation results to evaluate the performance of our proposed algorithm in estimating range and Doppler shft. Table \ref{tab:my_label} summarizes the key simulation parameters. The simulation parameters are set based on \cite{3GPPTR38901}. $c_1$ and $c_2$ can be set referring to \cite{AffineAli} to achieve the path resolvability. The effectiveness of our MPCPDP-based sensing method is influenced by several environmental factors, including the number of distinguishable MPCs, Signal-to-Noise Ratio (SNR) level, Nakagami-\(m\) distribution shape parameter \(m\), and target velocity \(v\). The following subsections analyze the impact of these factors on ranging accuracy. For each scenario, we conducted 1,000 simulations, computing the Root Mean Square Error (RMSE) for range estimation and the Normalized RMSE (NRMSE) for the Doppler shift \(\nu_1\) of the LoS path. Specifically, for \(m = 1\), we calculate the square root of the Cramer-Rao Bound (RCRB) to assess the precision of our sensing method. Additionally, we benchmark our results against the SoTA RSS-based ranging method, referred to as RSS-Nakagami in this paper, as described in \cite{AhmedSAS}. The thresholds \(\epsilon_1\) and \(\epsilon_2\) are set to \(10^{-3}\), and the maximum number of iterations for both the Expectation-Maximization and Expectation-Conditional algorithms is 1,000.

\subsection{Impact of SNR}
\begin{figure}
    \centering
    \includegraphics[width=0.9\linewidth]{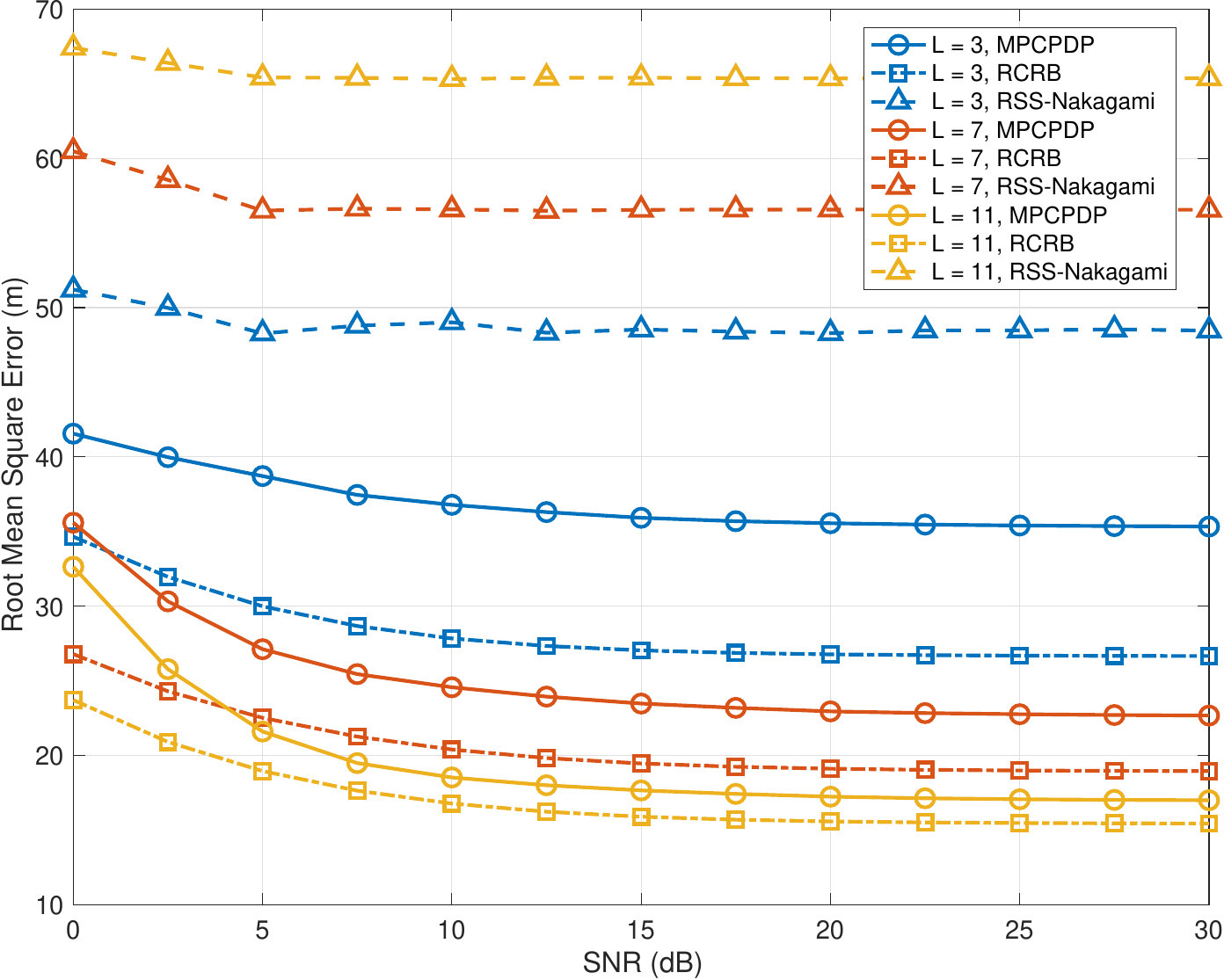}
    \caption{$m = 1$: RMSE of Range Estimation Across Different Paths and Methods vs. SNR}
    \label{fig:m1rangeSNR}
\end{figure}

\begin{figure}
    \centering
    \includegraphics[width=0.9\linewidth]{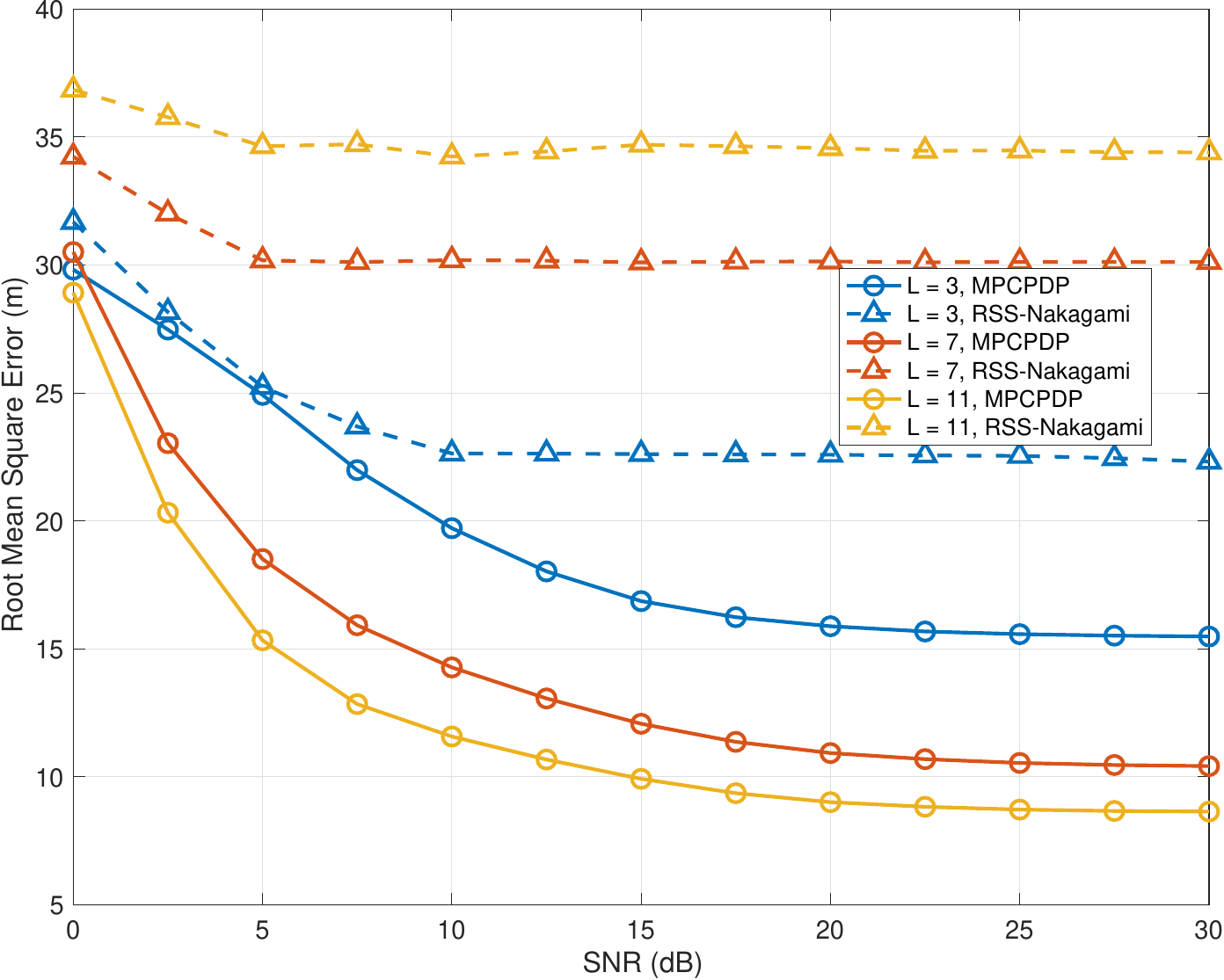}
    \caption{$m = 5$: RMSE of Range Estimation Across Different Paths and Methods vs. SNR}
    \label{fig:m5rangeSNR}
\end{figure}

In this series of experiments, we fixed the target velocity at 60 km/h to assess the impact of SNR on the range and Doppler shift estimation of the LoS path. The simulation results, depicted in Figs. \ref{fig:m1rangeSNR} and \ref{fig:m5rangeSNR}, compare ranging accuracy across different numbers of multipath components (MPCs) \(L\) with Nakagami shape parameters \(m = 1\) and \(m = 5\). The data reveal that increasing \(L\) improves ranging accuracy, as evidenced by the decreasing RMSE at all SNR levels. For instance, with an SNR of 10 dB, the RMSE values for three, seven, and eleven paths are approximately 36, 24, and 18 meters, respectively, reducing to about 9, 5, and 2 meters as the SNR rises to 30 dB. Additionally, an \(m\) value of 5 results in lower RMSEs, highlighting that reduced multipath fading enhances estimation performance. For example, at 10 dB SNR, RMSE values are around 20, 14, and 12 meters for three, seven, and eleven paths, respectively. These improve as the SNR increases to 30 dB. Our MPCPDP-based method outperforms the Nakagami-m based RSS-ranging approach, particularly in high-SNR settings. At \(m = 1\), the performance closely matches the theoretical lower bound (RCRB) at higher SNR levels, further validating the effectiveness of MPCPDP-based ranging.

\begin{figure}
    \centering
    \includegraphics[width=0.9\linewidth]{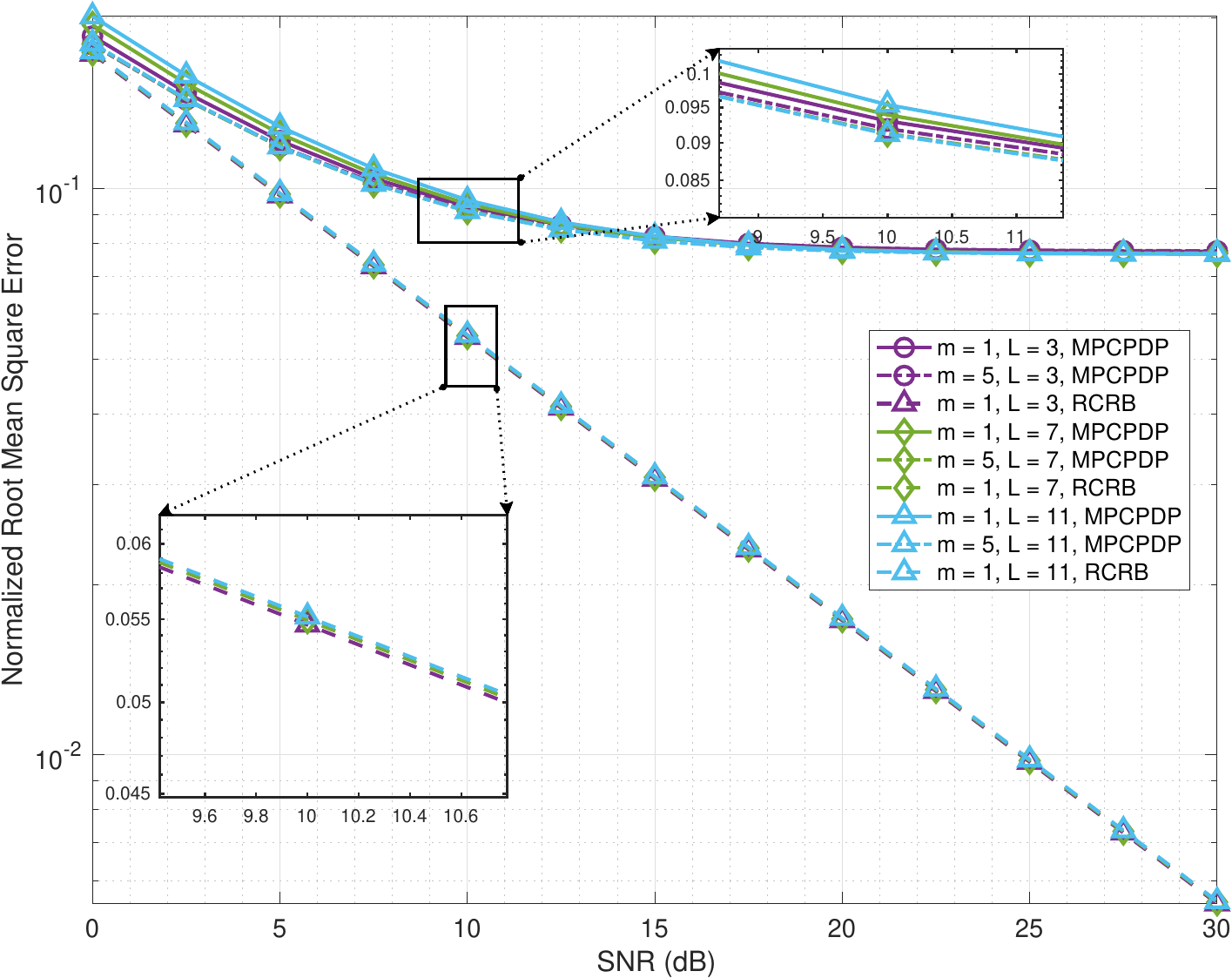}
    \caption{NRMSE of LoS Doppler Shift Estimation Across Different Paths and Nakagami-m Shape Parameters $m$ vs. SNR}
    \label{fig:dopplerSNR}
\end{figure}

Fig. \ref{fig:dopplerSNR} presents the Normalized RMSE (NRMSE) for Doppler shift estimations across different Nakagami shape parameters (\(m = 1\) and \(m = 5\)) and varied numbers of propagation paths (\(L = 3, 5, 7\)). These findings indicate that higher SNR improves accuracy, with NRMSE decreasing from approximately 0.2 to 0.07 as SNR increases from 0 dB to 30 dB. The number of paths shows minimal influence on accuracy, as the Normalized RCRB suggests. While minor discrepancies in estimation accuracy among different parameters exist, these are minimal compared to the effects of noise. The results also confirm the efficacy of the EM-EC algorithm, especially as estimation errors tend to diminish with increasing SNR.


\subsection{Impact of Absolute Velocity ($v$)}
\begin{figure}
    \centering
    \includegraphics[width=0.9\linewidth]{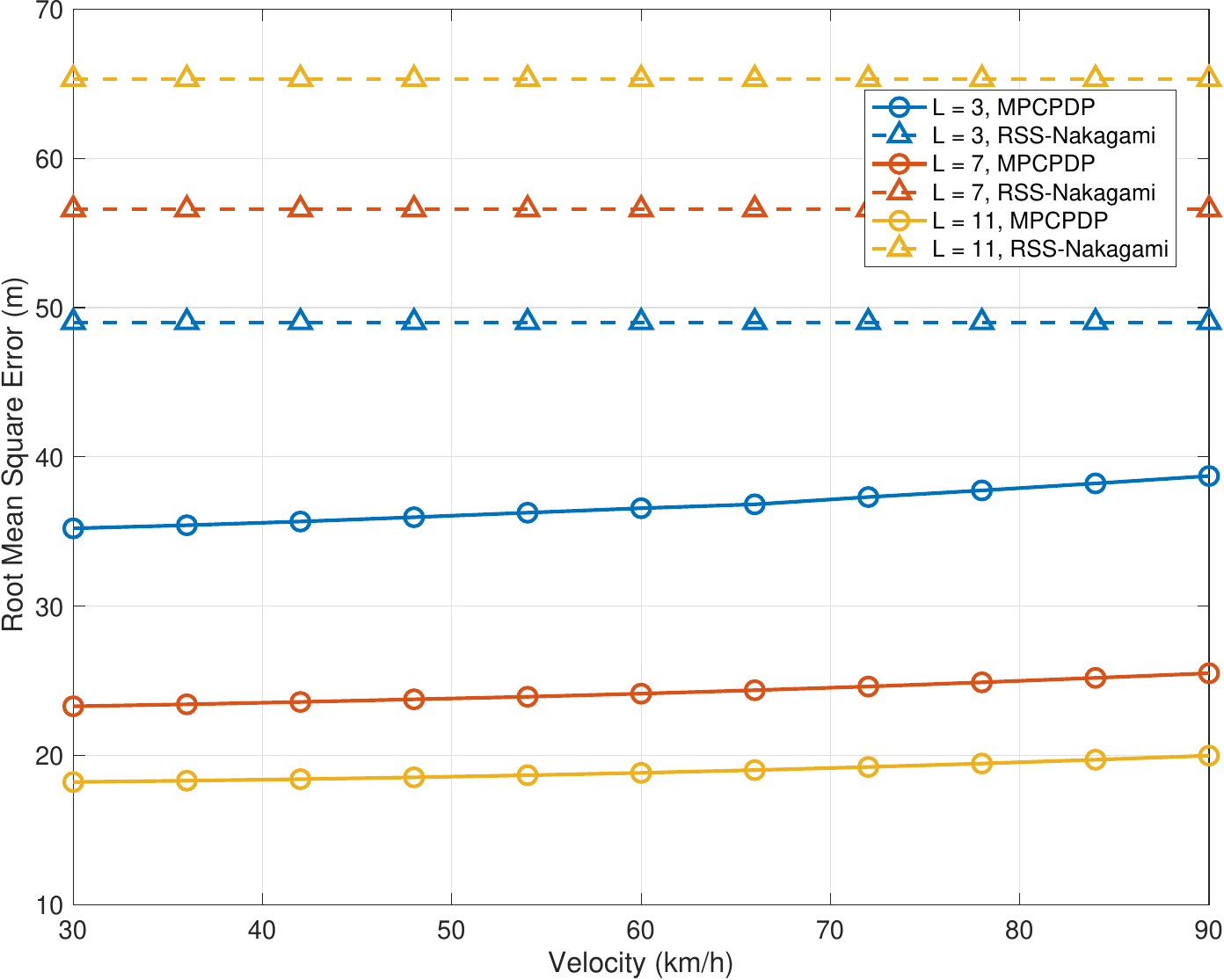}
    \caption{$m = 1$: RMSE of Range Estimation Across Different Paths and Methods vs. Velocity $v$}
    \label{fig:rangevelocitym1}
\end{figure}

\begin{figure}
    \centering
    \includegraphics[width=0.9\linewidth]{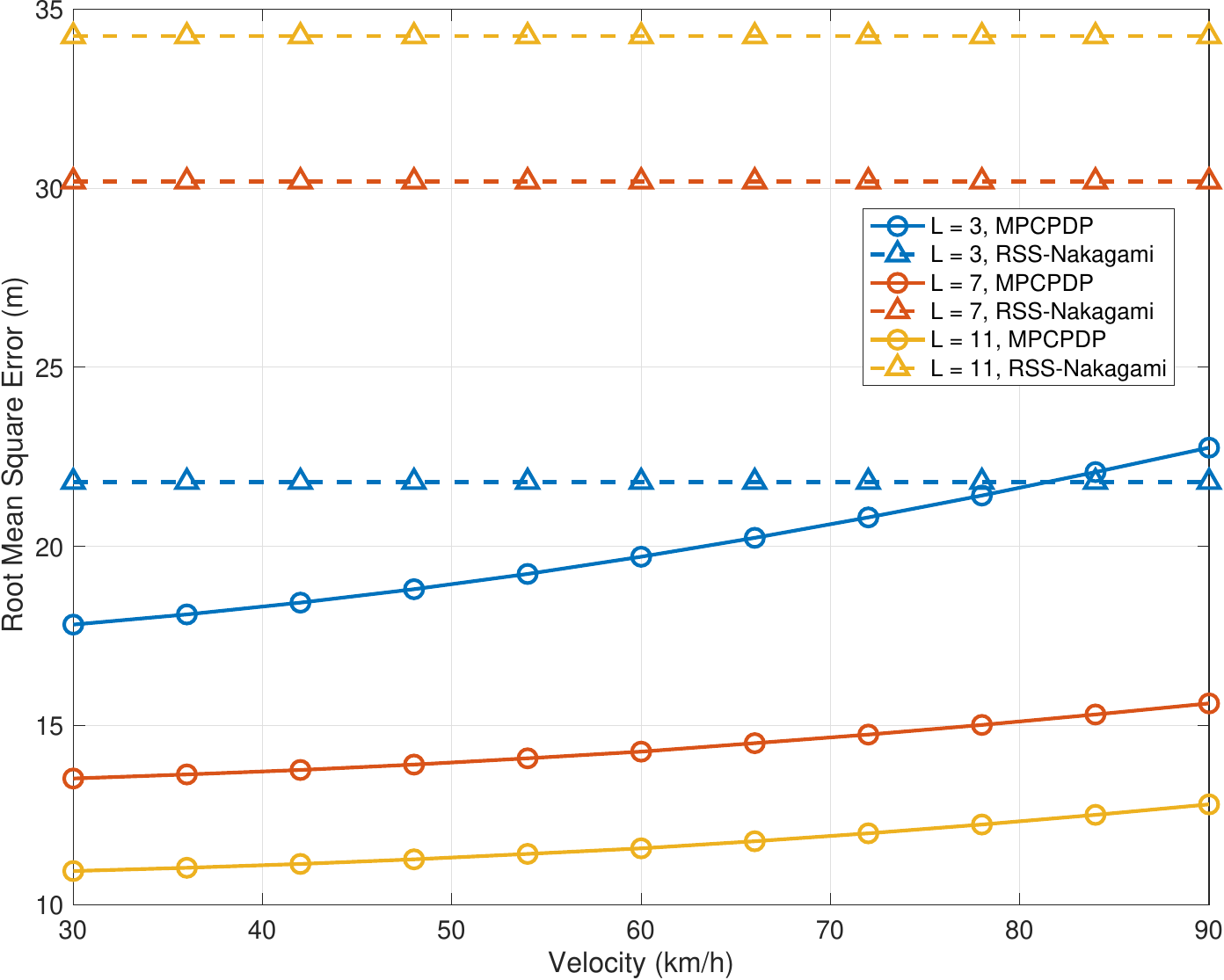}
    \caption{$m = 5$: RMSE of Range Estimation Across Different Paths and Methods vs. Velocity $v$}
    \label{fig:rangevelocitym5}
\end{figure}

In this experiment set, we fixed the SNR at 10 dB to assess its influence on range and Doppler shift of the LoS path estimation. Figs. \ref{fig:rangevelocitym1} and \ref{fig:rangevelocitym5} present the simulation outcomes for ranging with varying numbers of MPCs \(L\) under Nakagami shape parameters \(m = 1\) and \(m = 5\), respectively. This is due to the decreasing accuracy of the first-order Taylor expansion at higher Doppler frequency shifts. Although RSS-Nakagami method ranging slightly outperforms our method at a path count of three, \(m=5\), and speeds exceeding 80 km/h, the MPCPDP-based sensing method generally demonstrates superior accuracy in ranging, particularly as the number of paths increases. 

\begin{figure}
    \centering
    \includegraphics[width=0.9\linewidth]{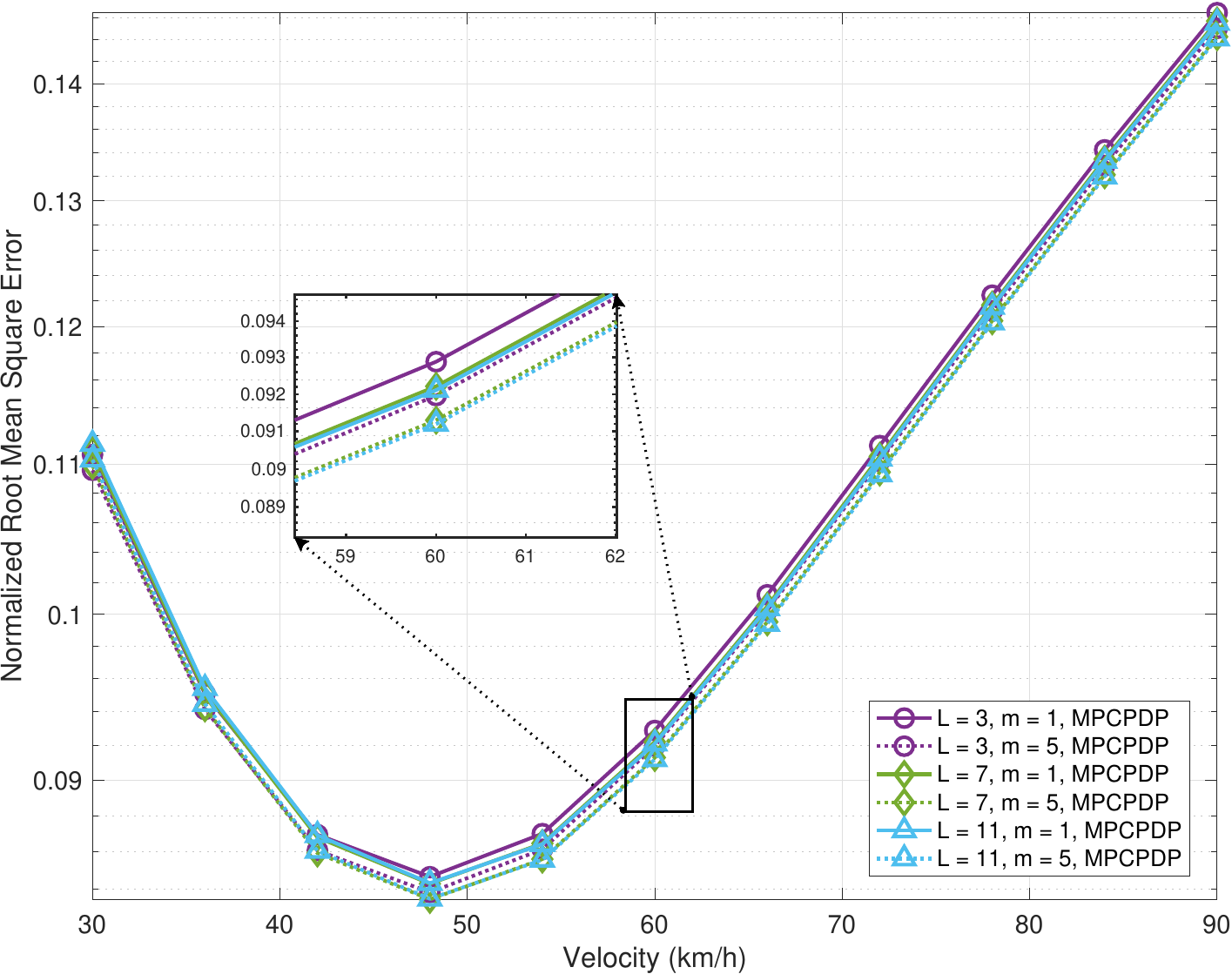}
    \caption{NRMSE of LoS Doppler Shift Estimation Across Different Paths and Nakagami-m Shape Parameters $m$ vs. velocity $v$}
    \label{fig:nuvelocity}
\end{figure}

Fig. \ref{fig:nuvelocity} shows the NRMSE performance of Doppler shift estimation of the LoS path across different Nakagami shape parameters (\(m = 1\) and \(m = 5\)) and varying numbers of propagation paths (\(L = 3, 5, 7\)), relative to changing velocity. From the perspective of the CRB for Doppler Shift, the magnitude of equal amplification Doppler shift does not compromise the theoretical accuracy of the estimated Doppler shift. At lower speeds, It is apparent that the NRMSE degrades slightly from 0.011 to 0.0087 as velocity increases from 30 km/h to 48 km/h for all cases, benefiting from the relatively stable error and high accuracy of the Taylor expansion. However, as velocity further increases, the accuracy of the first-order Taylor expansion around zero diminishes, potentially leading to poorer estimations.


\subsection{Impact of Shape Parameter \(m\) of the Nakagami-m Distribution}
\begin{figure}[!t]
	\centering
	\subfloat{
		 \includegraphics[width=0.48\linewidth]{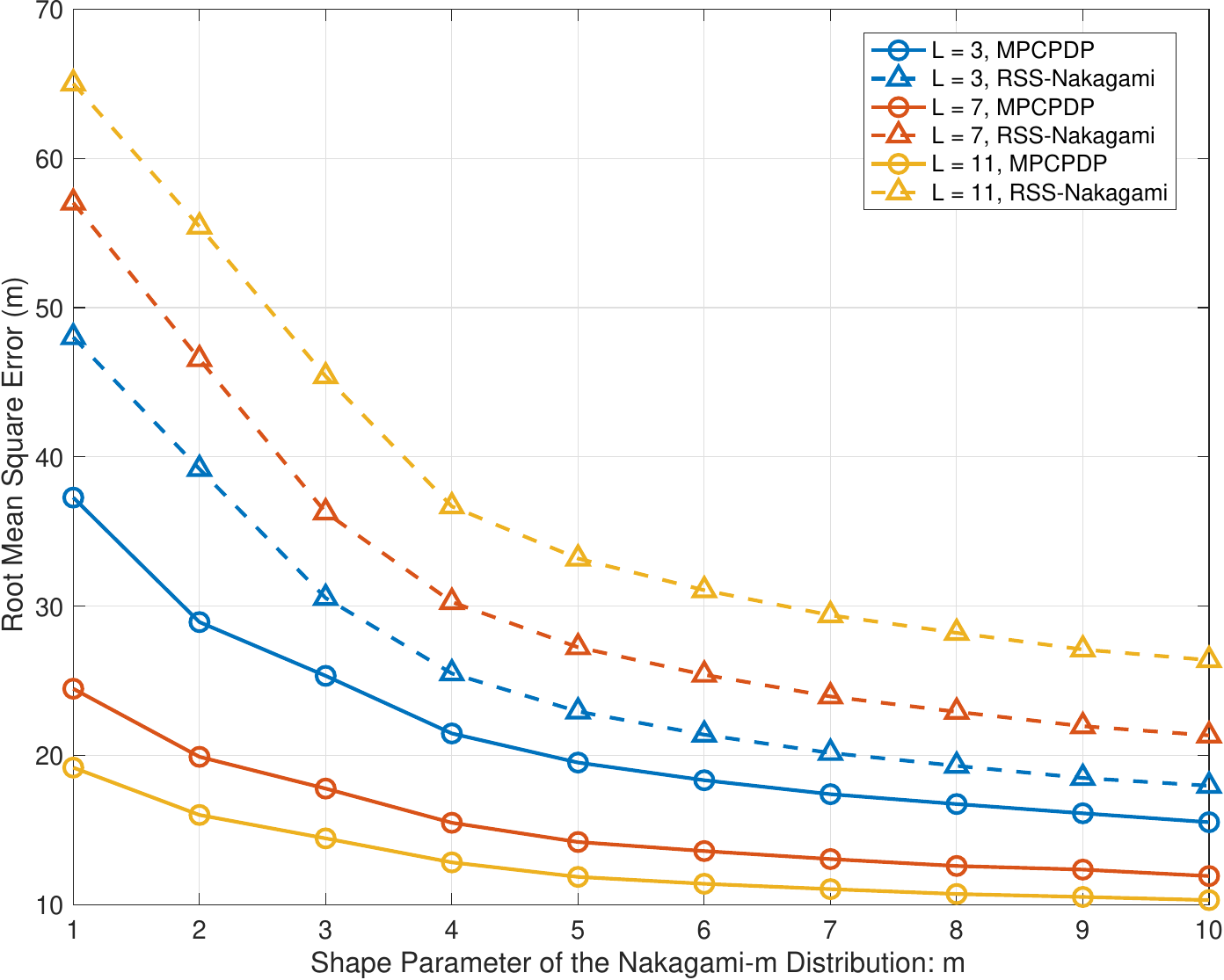}}
	\subfloat{
		\includegraphics[width=0.49\linewidth]{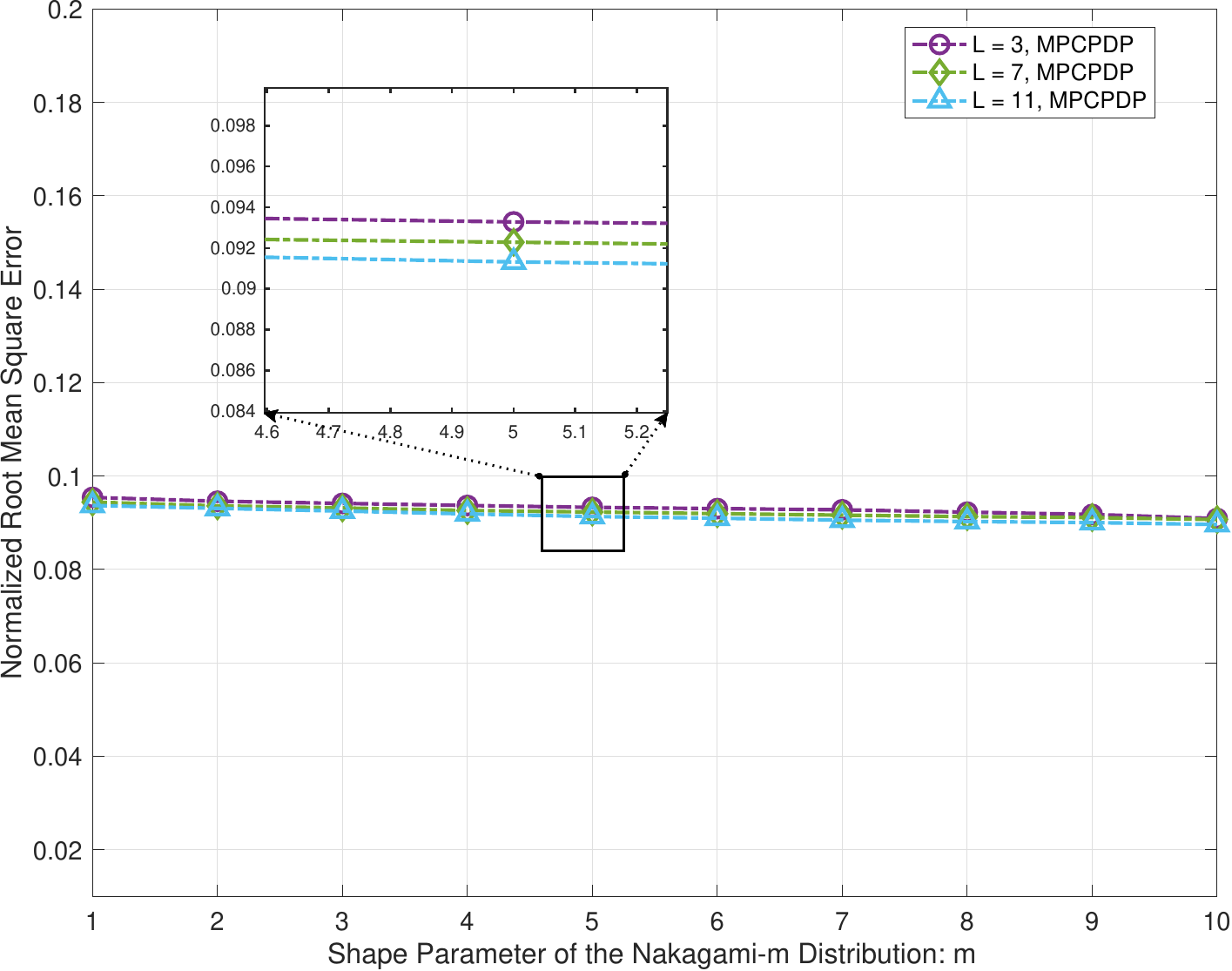}}
	\caption{Impact of the Shape Parameter $m$ For (a) Ranging and (b) Doppler Shift of the LoS Path ($v = 60$ km/h).}
	\label{fig:shapem}
\end{figure}

In this series of experiments, we fixed the SNR at 10 dB and the target velocity at 60 km/h to explore the effects of SNR on range and Doppler shift biases. Fig. \ref{fig:shapem} displays the impact on both ranging and Doppler shift estimations of the LoS path. Notably, ranging accuracy enhances with increasing \(m\) values, a result of channel energy converging more effectively as a function of propagation distance with higher \(m\) values. Regarding Doppler shift estimation, the accuracy is marginally influenced by the number of propagation paths, showing only about a 1\% difference, which is negligible compared to the impact of noise. Consequently, the shape parameter \(m\) has a minimal effect on the accuracy of Doppler shift estimation, indicating that noise is the dominant factor affecting these measurements.

\subsection{Impact of Missing Paths}
\begin{figure}[!t]
	\centering
	\subfloat{
		 \includegraphics[width=0.49\linewidth]{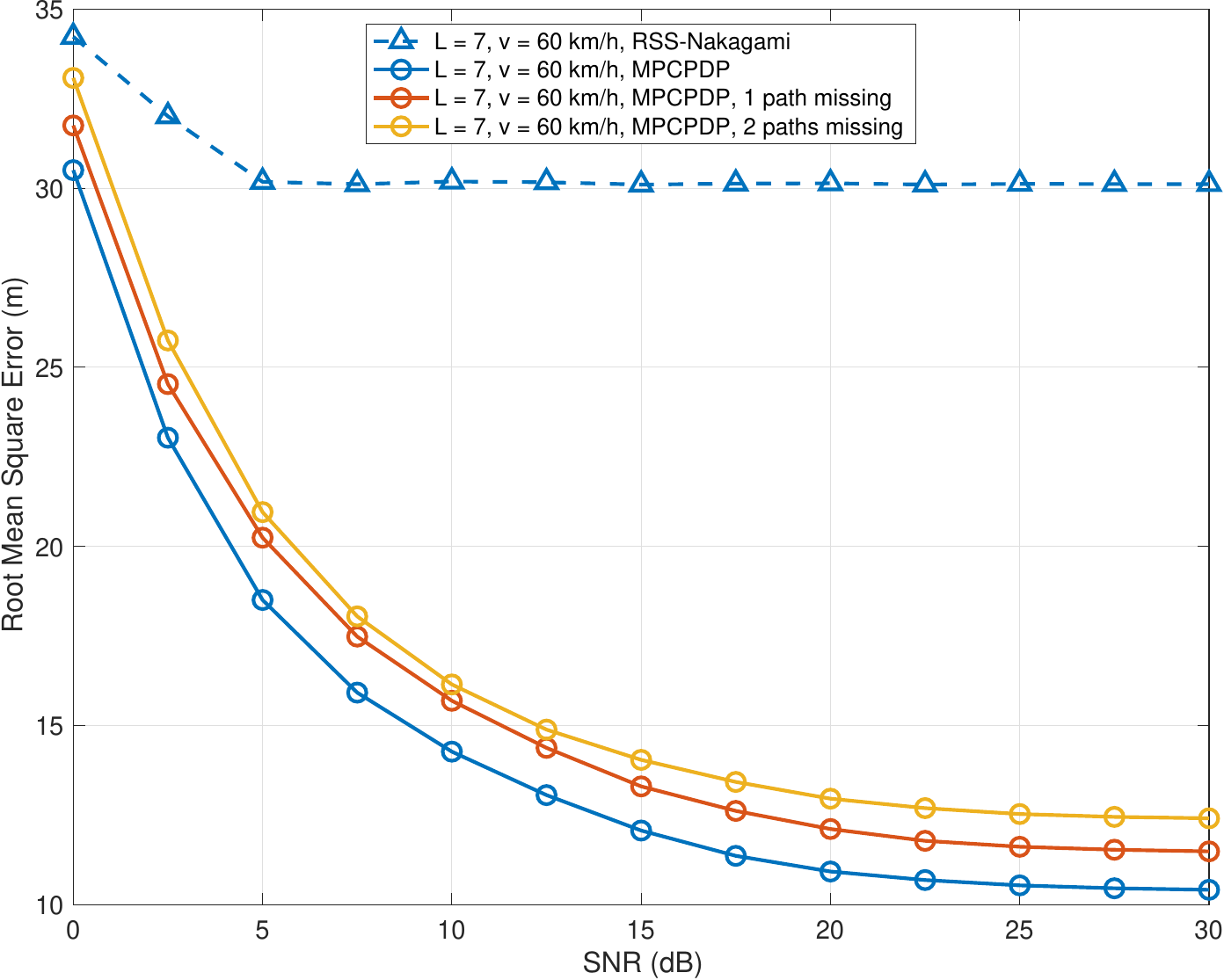}}
	\subfloat{
		\includegraphics[width=0.5\linewidth]{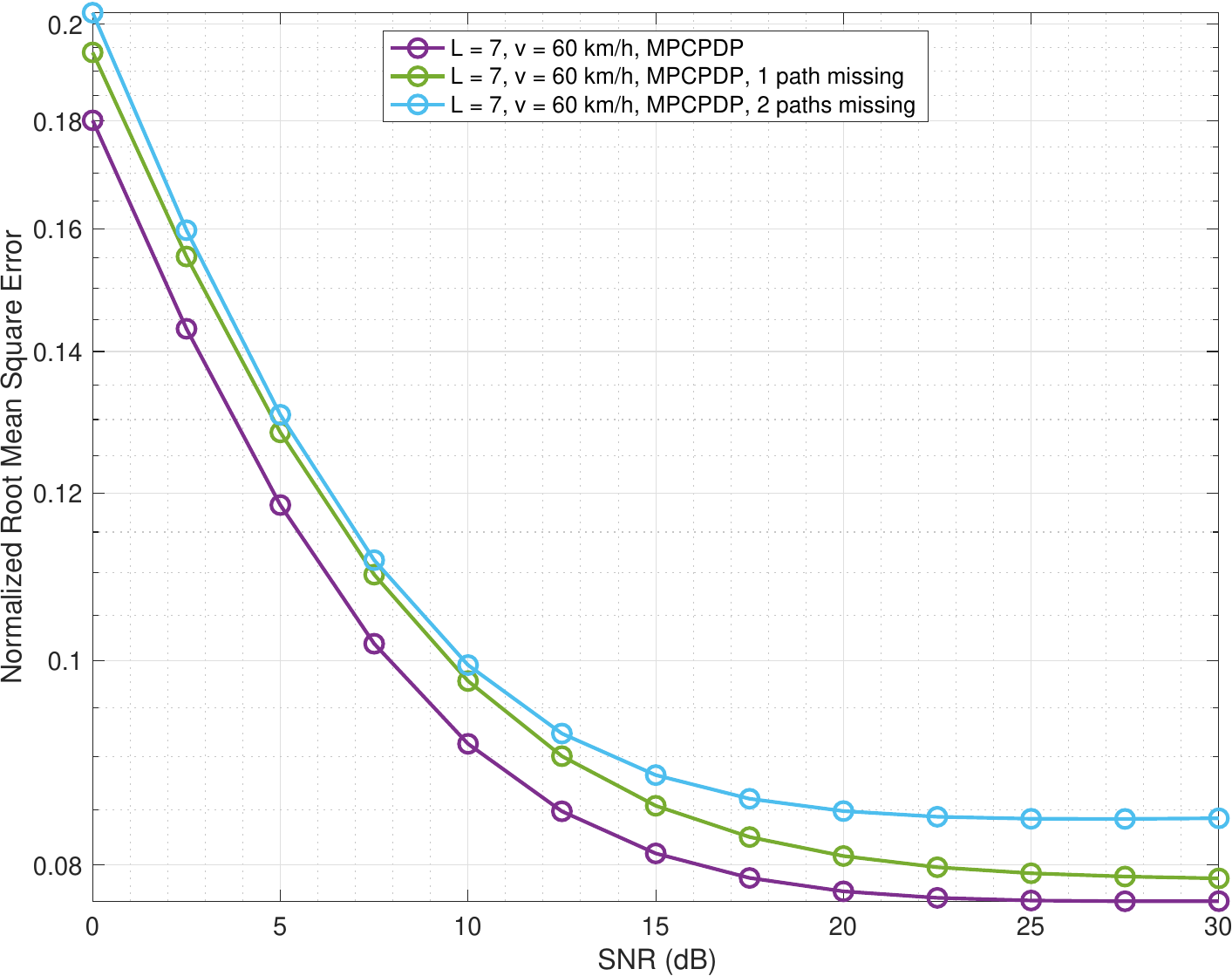}}
	\caption{$m = 5$: Impact of missing paths For (a) Ranging and (b) Doppler Shift of the LoS Path ($L=7, v = 60$ km/h).}
	\label{fig:Misspath}
\end{figure}

In this set of experiments, we used a fixed path count of seven and a Nakagami-m shape parameter \(m = 5\). Figure \ref{fig:Misspath} shows how missing paths during detection impacts the accuracy of ranging and Doppler shift estimates for the LoS path. Specifically, we compared scenarios where one and two NLoS paths were missed. Our results indicate a performance decline with missing paths: the ranging normalized mean square error (NMSE) worsened by approximately 5\% and 10\%, and the Doppler accuracy decreased by 4\% and 11\%, respectively. Despite these declines, performance with two omitted paths still surpasses that of the RSS-based method. For Doppler shift estimation, while the number of missed paths moderately impacts performance, it remains relatively unaffected by noise. These findings underscore the robustness and viability of our algorithm, even when some paths are not detected.

\subsection{Simulation Conclusions}
Based on experimental simulations varying SNR values, velocity $v$, shape parameter $m$, and the number of NLoS paths, our method has shown strong performance across diverse and complex environments. Notably, the number of NLoS paths and the shape parameter $m$ are critical factors, significantly influencing ranging estimation accuracy. As the number of NLoS paths increases, our MPCPDP-based sensing method improves in ranging accuracy, while Doppler shift estimation almost remains consistent. Additionally, an increase in $m$ reduces the variance of the fading channel, which enhances our method's performance. Although higher velocities slightly reduce the performance of our sensing method, the overall impact is minimal. These results underscore the effectiveness of our MPCPDP-based sensing in numerical simulations under the AFDM system, particularly when considering Doppler shifts.

\section{Conclusions} \label{secVII}
This paper investigates joint range and Doppler estimation in AFDM-ISAC systems, leveraging MPCPDPs. The unique properties of the AFDM system enable the resolution of LoS and NLoS paths within one pilot symbol, even in fast time-varying channels. We propose a novel ranging method that utilizes the range-dependent magnitude of MPCPDP across its delay spread, eliminating the need for additional time synchronization or extra hardware. To address the nonlinearities in Doppler estimation, a First-order Taylor expansion transforms the problem into a bilinear estimation challenge, which we address using the EM algorithm enhanced by the EC algorithm-termed EM-EC. Additionally, we derive the CRB for joint LoS range and Doppler shift estimation under Rayleigh fading, as assumed by a Nakagami-m distribution, and establish the fixed point for the EM-EC algorithm in other cases. Extensive simulations validate the effectiveness of our proposed method in ISAC systems, demonstrating promising results for enhancing the dual functionality of sensing and communication in wireless networks. To further validate the practicality and effectiveness of our method, our next step is to collect measurement data from diverse environments and conduct comprehensive experimental analyses. 
\bibliographystyle{IEEEtran}
\bibliography{MultipathComponentsPDPBasedSensing}
\end{document}